\documentclass[12pt]{article}

\usepackage{braket}
\usepackage{scicite}
\usepackage{graphicx}
\usepackage{times}
\usepackage{amsmath}
\usepackage{amsthm}
\usepackage{amsfonts}
\usepackage{bbm}
\usepackage{tabularx}
\usepackage{verbatim}
\usepackage{wrapfig}
\usepackage{caption}
\usepackage{subcaption}
\usepackage{float}
\usepackage{xcolor}
\usepackage{booktabs}

\usepackage[ruled,vlined]{algorithm2e}
\usepackage[retainorgcmds]{IEEEtrantools}

\newtheorem{theorem}{Theorem}
\newtheorem{assumption}{Assumption}

\usepackage[margin=.5in]{geometry} 

\newenvironment{sciabstract}{%
\begin{quote} \bf}
{\end{quote}}

\title{Interpreting convolutional neural networks' low dimensional approximation to quantum spin systems}

\author
{Yilong Ju,$^{1, \dagger}$ Shah Saad Alam,$^{2, \dagger}$ Jonathan Minoff$^{2}$, \\
Fabio Anselmi$^{3,4}$, Han Pu$^{2\ast}$, Ankit Patel$^{1,5\ast}$ \\
\\
\normalsize{$^{1}$Department of Computer Science, Rice University, 6100 Main St., Houston, TX 77005, USA}\\
\normalsize{$^{2}$Department of Physics and Astronomy, Rice University, 6100 Main St., Houston, TX 77005, USA}\\
\normalsize{$^{3}$Center for Neuroscience and Artificial Intelligence, Baylor College of Medicine,
Houston}
\\
\normalsize{$^{4}$Center for Brains, Minds, and Machines,
MIT, Cambridge, MA, USA}
\\
\normalsize{$^\dagger$Equal contribution.}
\\
\normalsize{$^\ast$To whom correspondence should be addressed; E-mail:  hpu@rice.edu, abp4@rice.edu}}

\date{}

\begin{document} 




\maketitle


\begin{sciabstract}
Convolutional neural networks (CNNs) have been employed along with Vari-
ational Monte Carlo methods for finding the ground state of quantum many-
body spin systems with great success. In order to do so, however, a CNN with only linearly many variational parameters has to circumvent the ``curse of dimensionality'' and successfully approximate a wavefunction on an exponentially large Hilbert space.
In our work, we provide a theoretical and experimental analysis of how the CNN optimizes learning
for spin systems, and investigate the CNN's low dimensional approximation. We first quantify the role played by physical symmetries of the underlying spin system during training. We incorporate our insights into a new training algorithm  and demonstrate its improved efficiency, accuracy and robustness. We then further investigate the CNN's ability to approximate wavefunctions by looking at the entanglement spectrum captured by the size of the convolutional filter. Our insights reveal the CNN to be an ansatz fundamentally centred around the occurrence statistics of $K$-motifs of the input strings.
We use this motivation to provide the shallow CNN ansatz with a unifying theoretical interpretation in terms of other well-known statistical and physical ansatzes such as the maximum entropy (MaxEnt) and entangled plaquette correlator product states (EP-CPS). Using regression analysis, we find further relationships between the CNN's approximations of the different motifs' expectation values. Our results allow us to gain a comprehensive, improved understanding of how CNNs successfully approximate quantum spin Hamiltonians and to use that understanding to improve CNN performance.

\end{sciabstract}

\tableofcontents

\section{Introduction}

The central concern of quantum many-body system is to understand how macroscopic properties emerge from microscopic inter-particle interactions. However, this is in general an extremely difficult question to answer due largely to the fact that the dimension of the quantum Hilbert space grows exponentially as the number of constituent particles increases. Ingenious numerical techniques have been developed to study certain classes of many-body systems. In recent years, techniques inspired by machine learning, specifically neural networks (NNs), have attracted much attention. In particular, Convolutional Neural Networks (CNNs), augmented with quantum Monte Carlo methods, have arisen recently as a potent class of variational ansatzes for numerically solving quantum spin systems with many particles \cite{Liang2021, miles2020correlator, roth2021group, liang2018solving}. CNNs  have often provided rapid and quite accurate numerical approximations, comparable to the traditional algorithms that exist in quantum physics. As a result, there has been a flurry of research to improve the performance of these  models and to apply them to broader classes of quantum spin systems with different physical constraints.  However, the exact approximations and methods used by the CNNs remain a mystery, with the CNNs effectively remaining mostly as black boxes. Indeed, this is a general problem for applications involving NNs, which has prevented us from being able to interpret the NN’s solution and to extract useful physical insights about quantum systems under study.  As a result, there is a lack of clear understanding on the full potential of machine learning on quantum research.

In this work, we take a crucial step in filling this gap. Specifically, we aim to give new insights into how even a simple, one-hidden-layer CNN provides a solution to a quantum spin problem. We show how physical features, such as symmetries of the quantum spin system, naturally manifest themselves in the final trained network and during the optimization dynamics. We show the constraints these symmetries place on the variational parameters, and we use these insights to construct a more efficient, accurate and robust training algorithm for CNNs. To further understand why the CNN is so adept at sufficiently approximating the system using linearly many parameters, we interpret the convolutional operation in terms of the degree of quantum entanglement captured by the CNN ansatz. Next, to interpret the advantages conferred by the mathematical form of the CNN, we provide a mapping of the CNN to other statistical and physical ansatzes such as Maximum Entropy (MaxEnt) and Correlator Product States (CPS). We also conduct a novel multivariate regression analysis to uncover which physical features are the most relevant to the low-dimensional learned solution and which ones the CNN captures correctly. Finally, we discuss how our approach and new insights can be used to design efficient approximations of complicated quantum spin systems.
\begin{figure}[h]
    \centering
    \includegraphics[width=\textwidth,trim={0 0 0 0},clip]{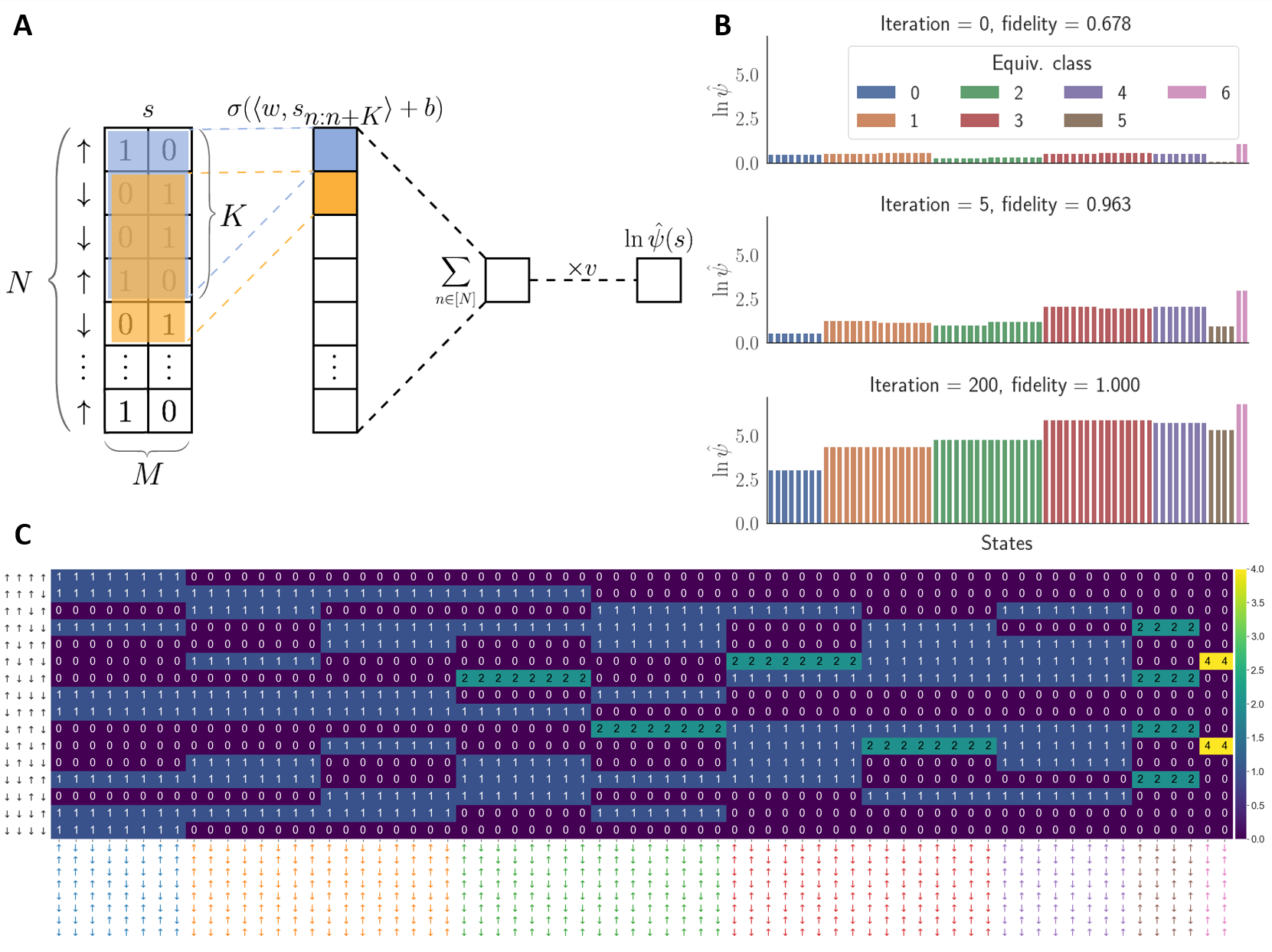}
    \caption{\textbf{A:} The CNN architecture for a system of $N$ sites with $M=2$ internal states. It has 1 filter and 1 convolutional layer with kernel size $K=4$, a ReLU activation function $\sigma(\cdot)$ and uses the cyclic padding. The one-hot encoding is used for each input state $s$ from the net $S^z=0$ manifold. \textbf{B:} Learned $\ln \psi(s)$ at different iterations.  \textbf{C:} Motif count matrix in the case $N=8$, $M=2$, and $K=4$. All $M^K = 2^4=16$ motifs are labeled for each row and all ${N \choose N/M} = {8 \choose 4} = 70$ states for each column. The counts are labeled with colors with larger motif counts having brighter colors. States are color-coded by the equivalence classes they belong to.} 
    \label{fig:CNN-arch}
\end{figure}
\section{The CNN Ansatz}
\subsection{CNN Architecture and Training}
For our choice of a physical toy system, we pick the 1-dimensional Sutherland model with periodic boundary conditions and Hamiltonian
\begin{equation}
    H=\sum_{i=1}^N P_{i,i+1}
\end{equation} where $P_{i,i+1}$ is the operator  exchanging the particles at positions $i$ and $i+1$, and the $N$ particles are evenly distributed among $M$ different species. For $M=2$, this system reduces to the antiferromagnetic spin-1/2 Heisenberg model. The reason we choose this Hamiltonian is twofold. First, it is simple enough that we can benchmark the CNN's solution by comparing its energy to the exact value given by the Bethe ansatz \cite{Sutherland1975}. Second, it is complex enough that the exact solution consists of $O(M^N)$ unique numbers, whereas the CNN only has $O(N)$ variational parameters to work with. In order to succeed, the CNN must find a way to efficiently represent an approximation to the exact solution, and we seek to understand the nature of this approximation.

To investigate the physics as simply as possible, we start with a basic CNN with a single convolutional layer followed by a fully connected layer (see Fig.~\ref{fig:CNN-arch}A). The inputs to this CNN are the spin configurations $s=\{s_1,s_2,...,s_N\}$ and output is $\ln \psi(s)$, where $\psi(s)$ is the wavefunction at $s$ parametrized as:
\begin{equation}
    \ln \psi^\mathrm{CNN}(s) = v \sum_{i=1}^N \sigma ( w \cdot s_{i:i+K-1}+b), \forall s \in \mathcal{S}_{N, M}, \label{eq:CNN_arch}
\end{equation}
where $\sigma$ is the ReLU non-linearity, $w \in \mathbb{R}^K$ is a convolutional filter of size $K$, $b \in \mathbb{R}$ is a scalar bias, $v \in \mathbb{R}$ is a scalar weight, and $s_{i:i+K-1}$ is the substring of $s$ of length $K$ starting at index $i$.  Since the Sutherland model does not allow for changes in total magnetization, we have restricted our input spin configurations $s$ to have zero net magnetization i.e. $s \in \mathcal{S}_{M,N}$. We note in passing that, for this particular problem, a nonlinear activation function is required for preventing the CNN from producing constant outputs (see Sec. \ref{ap:sec:failure} for proof).

Interestingly, if we combine the training results reported in Fig.~\ref{fig:CNN-arch}B with the strings shown in Fig.~\ref{fig:CNN-arch}C which have the same color as the bars in Fig.~\ref{fig:CNN-arch}B, we can see that a pattern emerges: certain strings $s$ have very similar $\ln\psi(s)$ to each other. On further inspection, we see that the states that have similar $\ln\psi(s)$ are the ones that are connected to each other by a combination of symmetry operations of the Hamiltonian: translations, reflections around any point, and permutations of the spin labels. Essentially, the CNN efficiently captures the symmetry constraints of the target function after training. Our goal is to see how these symmetries in the target function manifest within the CNN's variational parameters itself.

As mentioned earlier, the CNN cannot directly `see' the full input string $s$ of size $N$; instead it gleans information about $s$ indirectly through substrings $s'$ of size $K$ that it can `see' directly via the convolution operation. We call these substrings $K$-motifs. In order to learn about the global symmetries of the Hamiltonian, the CNN must somehow glean this information using only the frequency and occurrences of the $K$-motifs, which we can visualize via a \emph{motif count matrix} shown in Fig.~\ref{fig:CNN-arch}C (see Sec. \ref{appendix:critical-kernel-size} for a mathematical definition). As we will see later, motifs are the key to understanding why a low-dimensional approximation to the ground state exists, and why the CNN is particularly suited for this task. Before giving a detailed explanation, we first turn our attention to how the symmetries of the problem appear within the CNN.

\subsection{Symmetries Reduce the Complexity of Ground State Wavefunction} \label{sec:symmetries}
In our quest to understand the CNN's approximation, we start looking into the role of symmetries in decreasing the complexity of the target ground-state wavefunction. The Sutherland Hamiltonian is invariant under three symmetries that are commonly found in physics: translation, reflection, and SU($M$) rotations among the $M$ types of particles. Let $\mathcal{G}$ denote the symmetry group generated by all of these symmetries. It follows that the unique, nondegenerate ground state must also obey these symmetries in $\mathcal{G}$. In the uncoupled spin basis, the positive definiteness of the wavefunction allows the SU($M$) symmetry to be reduced to an $\mathrm{S}_M$ symmetry defined by simply permuting the $M$ different particle labels.  Due to these symmetries, the target function the CNN must learn has unique values only in a quotient space, a subspace of the ambient Hilbert space, since $\psi_{GS}(gs)=\psi_{GS}(s)$, for all $g \in \mathcal{G}$. The symmetries partition the Hilbert space $\mathcal{H}$ into equivalence classes of symmetric states $\mathcal{E} \equiv \mathcal{H} \mod \mathcal{G}$. The number of equivalence classes $|\mathcal{E}|$ can be computed exactly for small $N$, and for large $N$, a lower bound is given by $|\mathcal{E}| \geq \frac{N!}{2NM!\left(\left(N/M\right)!\right)^M}\sim O(M^N)$ (see Supplement \ref{supp:ec}). We thus find that the symmetries reduce the complexity of the target function the CNN is required to learn, but its complexity is still exponential in $N$ assuming a constant number of parameters needed per equivalence class. 

\subsection{Representing and Approximating Symmetries with CNNs} \label{sec:symmetries-approx}
\begin{figure}
    \centering
    \includegraphics[width=.45\textwidth]{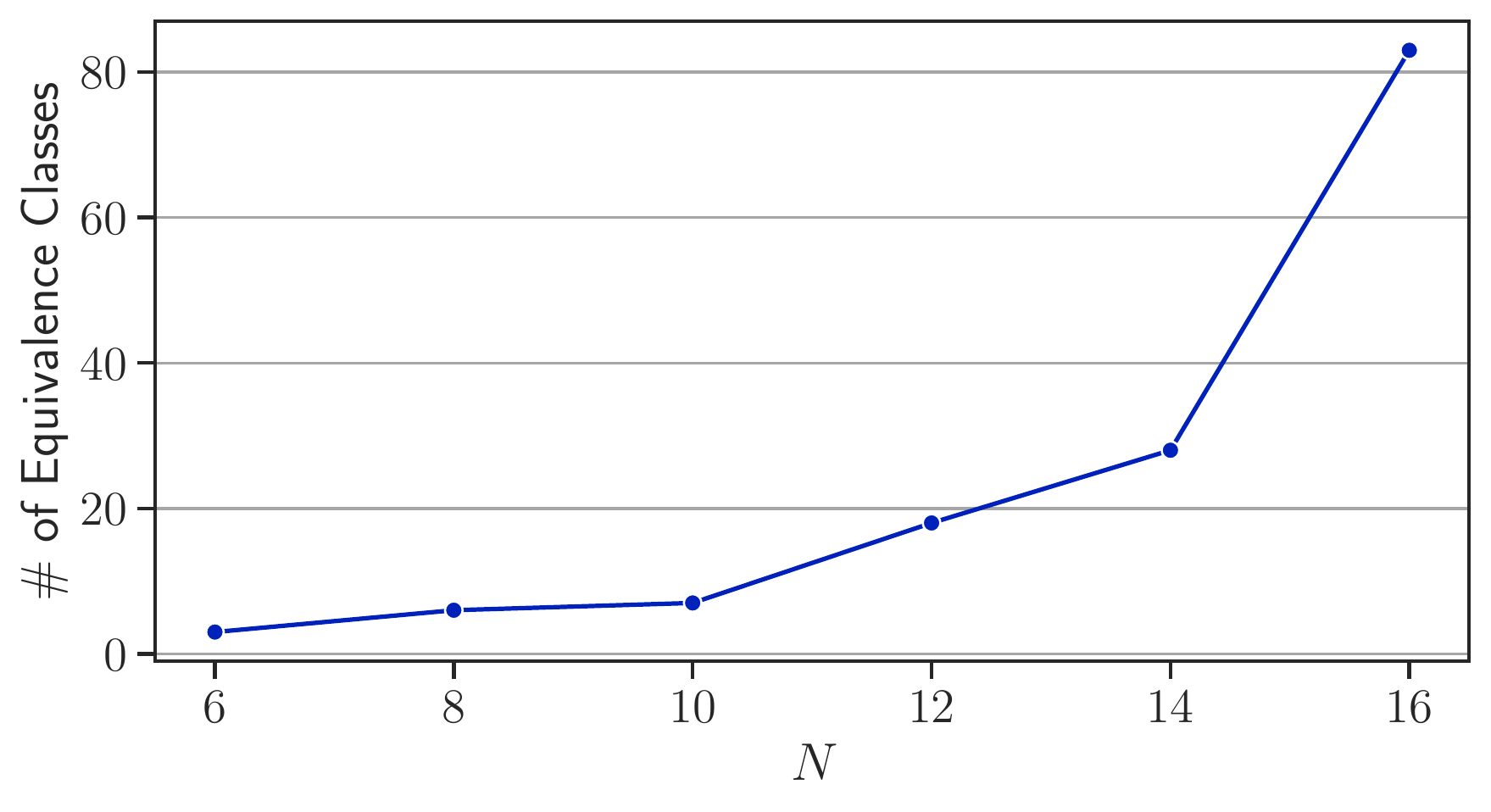}
    \includegraphics[width=.45\textwidth]{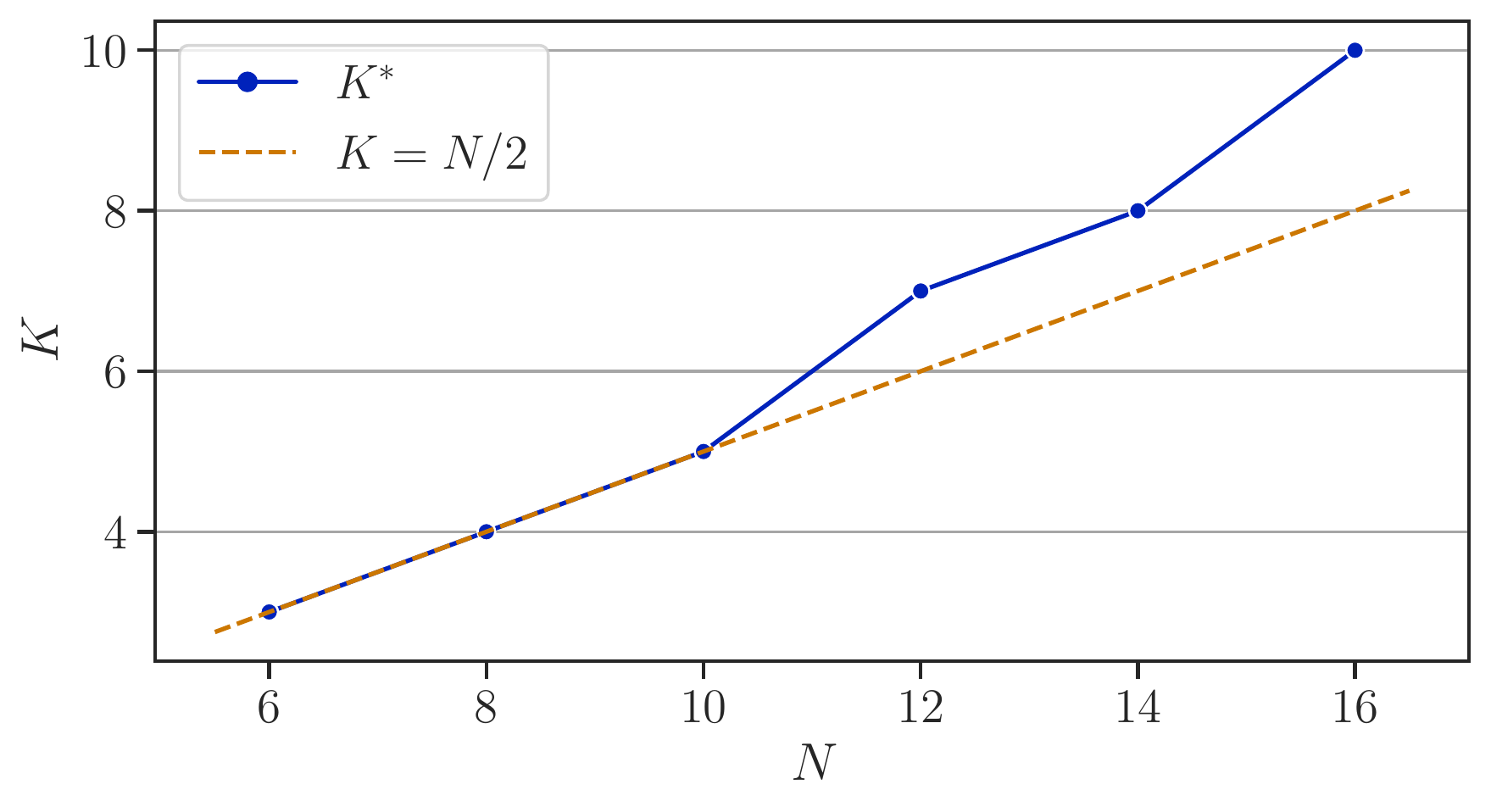}
    \caption{\textbf{Left:} Number of equivalence classes required to form a 99\% accurate approximation of the exact ground state wavefunction. The exponential scaling of this quantity demonstrates that finding the right approximation is not a trivial task. \textbf{Right:}  Critical kernel size $K^*$ vs. $N$ in the case of $M=2$. We empirically observe a growth rate slightly faster than $O(N)$.}
    \label{fig:classcutoff}
\end{figure}

And yet the success of CNNs is proof that a more parsimonious approximation does indeed exist. Might the CNN be learning by only selectively approximating some more important equivalence classes, or perhaps by taking advantage of strong dependencies between equivalence classes? While the exact ground state wavefunction consists of exponentially many amplitudes, a small subset of equivalence classes might account for the majority of the probability mass. To test this, we computed 
the minimum number of equivalence classes required to achieve a 99\% of the cumulative ground state probability as a function of system size $N$. Fig.~\ref{fig:classcutoff} (Left) shows that this still scales exponentially with $N$, implying that complexity reduction due to symmetry constraints cannot fully explain how a CNN can achieve high accuracy with only polynomially many $O(K)$ variational parameters. Furthermore it is not immediately clear why this dimensionality reduction is even physically possible. We will revisit this issue later on.

We next turn to the question of how the CNN learns and represents the symmetries of the Sutherland Hamiltonian. The equivalence classes are the result of global symmetries of the Hamiltonian which are only manifest in the full spin configuration or string of length $N$. However, as we saw before, the CNN cannot 'see' the full length $N$ string, but rather can only `see' substrings of length $K$ via a convolutional filter of size $K<N$. How large must $K$ be in order to learn an accurate approximation? Supp. Sec. \ref{appendix:critical-kernel-size} shows that in order to distinguish all equivalence classes we need filters of size at least $K>N/3$. 

However, the key insight lies in the motif count matrix. The rank of this matrix exactly equals the number of basis states that the CNN is able to distinguish, so in order to differentiate all equivalence classes, the rank of the motif count matrix must be at least as large as the number of equivalence classes. We define $K^*$ as the minimum value of $K$ that satisfies such condition. We show its growth vs. $N$ in Fig. 2 (Right). For $M=2$ and $K<N/2$, the rank of the motif count matrix is equal to $2^{K-1}$, which establishes a connection between the equivalence classes and the convolutional operation. However, for sufficiently large $K$, the CNN fails to recognize reflection or permutation symmetry, and instead partitions the states only into translationally-invariant subclasses of each equivalence class. 

\paragraph{Representing Symmetry: the Grand Sum Condition.} How do symmetry constraints manifest themselves in the trained CNN's variational parameters? We give a crucial part of the answer in the following theorem (see Supp. Sec.~\ref{supp:sec:gs} for the proof):
\begin{theorem} \label{thm:gs}
    For systems with $M = 2$, if $\mathrm{grandsum}(w) + 2b = 0$, then the CNN wavefunction possesses the relabeling symmetry.
\end{theorem}

Note that this condition does not directly insure reflection symmetry. But for strings in certain equivalence classes (eg. the orange and red classes in Fig.~\ref{fig:CNN-arch}C), applying a reflection is equivalent to first applying a relabeling operation followed by a proper translation. Thus, imposing the Grandsum condition ensures that strings in these equivalence classes will also possess the reflection symmetry. 

\subsection{Improving CNN Performance by Imposing Symmetry Constraints}
This motivates us to find a way of imposing these symmetry constraints into the CNN. We next show that, by enforcing the grand sum condition in various ways, we can improve the CNN's accuracy, robustness to initial conditions, and training speed. We propose two symmetry-forcing algorithms: SymForce-Init, which enforces the grand sum condition only at initialization, and SymForce-Traj, which enforces the grand sum condition throughout the entire learning trajactory (i.e., after each parameter update at every iteration, see Sec.~\ref{supp:sec:alg} for details). Both are simple to implement and compute, and compatible with any training scheme, since calculating the grand sum is just summing over $2M + 1$ parameters (with typically $M \leq 5$ and $K \leq 100$). Our heuristic for the parameter projection is to set $v \leftarrow v$, $w \leftarrow w - (\mathrm{grandsum}(w) + 2b) / (2K)$ and $b \leftarrow b$ after updating $(v,w,b)$ at each iteration. In addition, we can prove that (see Sec.
 \ref{suppsec:learning_dynamics}) once the CNN has learned the symmetries, the update of $\psi(s)$ equals the update of $\psi(gs)$ for any state $s$ and transformations $g$ of interest. Then, it will not forget them during the rest of the training. Thus, we can expect SymForce-Init to have a similar performance to SymForce-Traj.

\paragraph{Experiment Setting.} 
To test these algorithms, we adapt the deep architecture used in \cite{yang_deep_2020} to its shallow version and train a 1-layer 1-filter CNNs using SymForce-Init and SymForce-Traj. We focus on very large $SU(2)$ systems, where $N \in \{60, 240\}$ and $M=2$. We use both the shallow (only 1 convolution layer) and the deep CNNs as the baselines, labeled as Original and Deep ($L$ layers), where $L$ stands for the number of convolutional layers. See Sec. \ref{suppsec:hyperparameters} for more hyperparameter settings and tuning details.


\begin{figure}[ht]
     \centering
     \includegraphics[width=\textwidth, trim={0 0 0 0},clip]{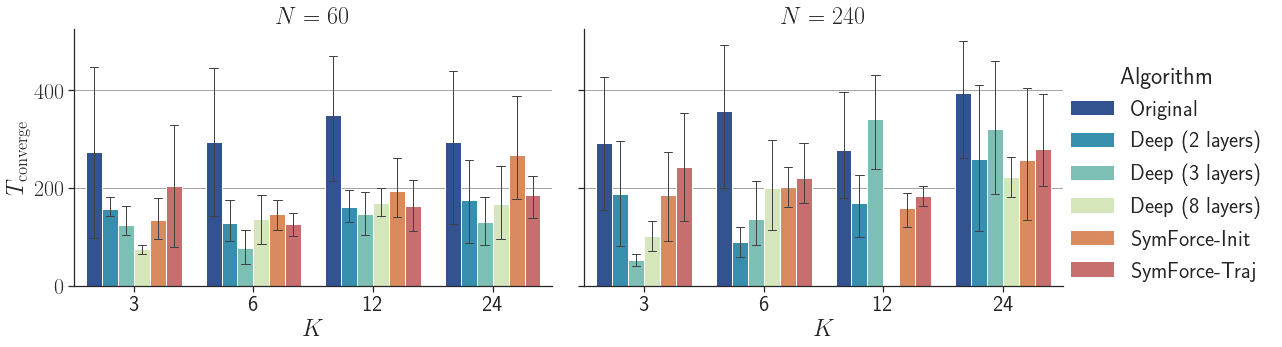}
     \includegraphics[width=\textwidth, trim={0 0 -170 0},clip]{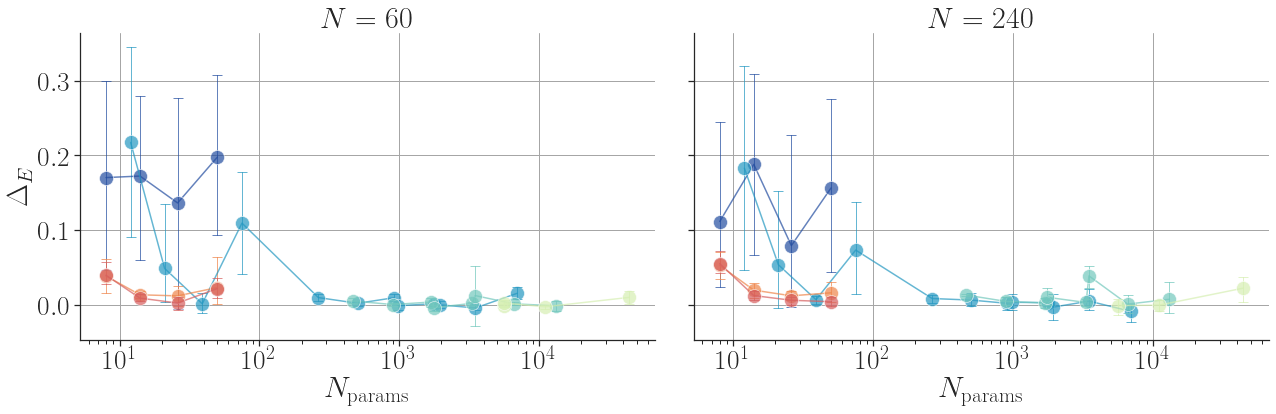}
     \caption{\textbf{Top:} The number of iterations until convergence $T_{\mathrm{convergence}}$ vs. kernel size $K$ for $N = 60$ and $N = 240$. \textbf{Bottom:} The error between the predicted and actual ground state energy $\Delta_E = \hat{E}_0 - E_0$ vs. the number of parameters in a CNN $N_{\mathrm{params}}$ (in log scale) for $N = 60$ and $N = 240$. We select the hyperparameters for each algorithm and $K \in \{3, 6, 12, 24\}$ corresponding to the minimum $\delta_E$, averaged over 5 random initializations. We use the same color map for the algorithms in both panels. We omit the bar for $N=240$, $K=12$ since there is no hyperparameter combination that ensures the training converges.}
     \label{fig:E-hat-alg}
\end{figure}

We monitor the number of iterations until convergence $T_{\mathrm{convergence}}$ and the error between the predicted and actual ground state energy $\Delta_E = \hat{E}_0 - E_0$ for each experiment setting, averaged over 5 random initializations. $T_{\mathrm{convergence}}$ is defined as the first iteration that the relative change of the rolling average of $\hat{E}_0$ compared to that at 5 iterations ago is smaller than 0.01\%. Since we only train the models for 500 iterations, $T_{\mathrm{convergence}}$ is set to 500 if this criterion is never met.

Fig. \ref{fig:E-hat-alg} shows that for both $N$ values, the proposed algorithms with only simple modification can indeed improve the CNN training. The top panel shows that compared to the vanilla training algorithm, CNNs trained using the symmetry-forcing algorithms can achieve a roughly $1/3$ reduction in $T_{\mathrm{convergence}}$, comparable to deeper CNNs. The bottom panel shows that with our symmetry-forcing algorithms, the shallow CNNs can achieve the same level of accuracy as the deeper CNNs even only using orders of magnitude fewer parameters, while still being robust to initial conditions and choice of hyperparameters. CNNs with 2 hidden layers seem to have similar advantages, but they are much more sensitive to the choice of hyperparameters, as we can see from the large error bars and the fact that only one $K$ works when $N_{\mathrm{params}} < 10^2$. Also, it is more difficult to interpret 2-layer CNNs, since the 2nd layer aggregates the motif activations and learns higher-level concepts. We also find that it does not always help by increasing the depth, in terms of both $T_{\mathrm{convergence}}$ and $\Delta_E$. In addition, we observe that SymForce-Init and SymForce-Traj have similar behavior in all aspects. This validates our statement regarding the learning dynamics.
\section{When does the Ground State Admit Parsimonious Approximations?}\label{sec:entanglement}
So far, we've studied how symmetry induces exact simplifications in the wavefunction and explored conditions for when the CNN successfully learns them. However, as mentioned in section \ref{sec:symmetries} symmetries are not enough to explain why a CNN architecture efficiently \emph{approximates} the wavefunction, sacrificing exactness.

To tackle this question we start analyzing the key operation of CNNs: the convolution. From the definition of the CNN wavefunction in Eq.~\eqref{eq:CNN_arch}, we can see that each term in the sum involves a single window of size $K$. The convolutional operation can thus be understood as the one that looks at positions $i:i+K-1$ for the occurrence of particular $K$-sized motif. In fact, we can rewrite the CNN in terms of a motif counting function $m^i_{s'}(s)$ that checks whether $s$ contains the motif $s'$ at position $i$. This allows us to rewrite the CNN as (see Sec. \ref{supp:lnpsi-CNN-string} for derivation):
\begin{equation}
    \psi^{\text{CNN}}(s)=\prod_{s' \in \mathcal{M}_{MK}} \exp\Big[ v \sigma(ws'+b)m_{s'}(s)  \Big]\equiv\prod_{s' \in \mathcal{M}_{MK}} \phi(s')^{m_{s'}(s) } \label{eq:CNNphi}
\end{equation}
where $\mathcal{M}_{MK}$ is the set of all possible motifs of size $K$ for $M$ particle types, $\phi(s')=\exp\left[v\sigma(ws'+b)\right]$ are free parameters, and  $m_{s'}(s)=\sum_i m^i_{s'}(s)$ is the number of times that motif $s'$ occurs in $s$ and forms the entries of the motif count matrix.
Note that the CNN is both a \emph{product} and \emph{exponential family} ansatz.

Written in this form, it's immediately clear that the CNN is never solving exactly on the full $M^N$ space: it is merely solving for a function $\phi(s')\in M^K$ and uses the motif frequencies to construct an effective product approximation for the full Hilbert space. However, $\phi(s')$ still has $O(M^K)$ entries, and the CNN has only $O(K)$ parameters and is therefore cutting corners even in learning $\phi(s')$. To gain insight into why and when the CNN can get away with doing this, we consider the role of entanglement in the CNN. While other studies have examined connections between entanglement and  Restricted Boltzmann Machine neural network states by working directly with the neural network state \cite{deng2017quantum, harney2020entanglement, sun2022entanglement}, we instead consider the reduced density matrix for $K$ adjacent particles.

The reduced density matrix $\rho_K$ for this $K$ spin subsystem can be calculated by first constructing the full density matrix $\rho=\ket{\Psi_{GS}}\bra{\Psi_{GS}}$ and then tracing out remaining $N-K$ spins. The expected frequencies of the motif counting operator $m_{s'}$ in the ground state can be found by  $\langle m_{s'}\rangle_{GS}= \langle s'|\rho_K|s'\rangle$. Although these observables would be easily obtainable experimentally, they can be computed exactly only for small systems via diagonalization. From the $\rho_{K}$ diagonalization, $\rho_K=\sum_\alpha e^{-\epsilon_\alpha}\ket{\alpha}\bra{\alpha}$, we obtain  $\langle m_{s'}\rangle=\sum_\alpha e^{-\epsilon_a}|\braket{s'|\alpha}|^2$, where $\{\ket{\alpha}\}$ and $\{e^{-\epsilon_\alpha}\}$ are the eigenvectors and eigenvalues of $\rho_K$ and the spectrum $\{\epsilon_\alpha\}$ is known as the \emph{entanglement spectrum} \cite{li2008entanglement}.

\begin{wrapfigure}{r}{0.5\textwidth}
    \centering
\includegraphics[width=0.48\textwidth]{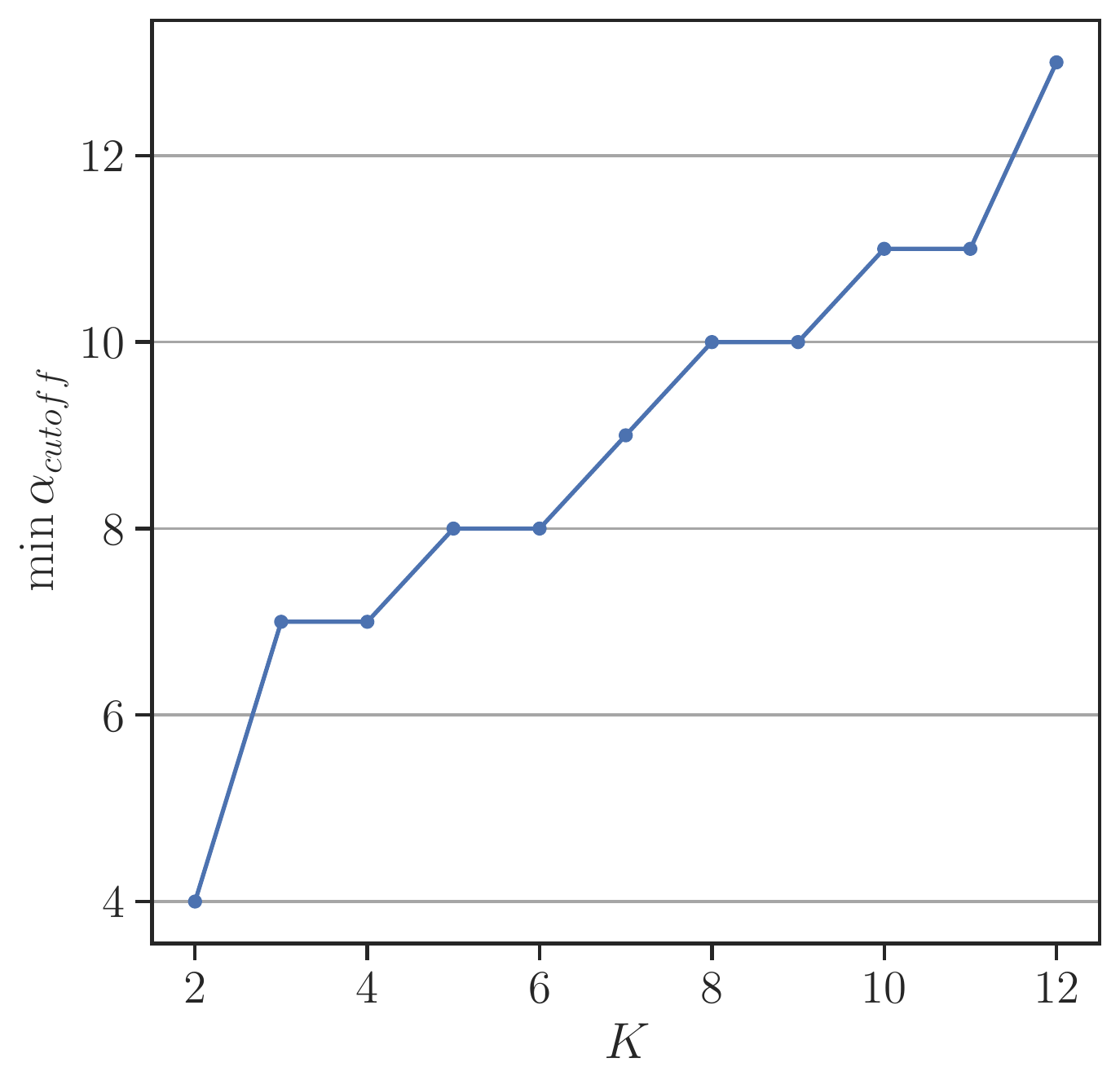}
    \caption{Size of the minimum truncation of $Z=\sum_\alpha e^{-\epsilon_\alpha}$ required to achieve 99\% accuracy, as a function of $K$. \vspace{-0pt}}
    \label{fig:alphacutoff}
\end{wrapfigure}

It’s immediately obvious that the CNN’s ability to use a motif to glean information about the entire system from just $K$ spins is related to the question of how “entangled” the $K$ substring is with the rest of the system. This reliance on quantum entanglement of a subsystem and its connection to the simulability of a particular quantum Hamiltonian is a well-explored concept in Matrix Product States (MPS), the dominant class of variational ansatzes used in solving 1D problems. The ability of the MPS ansatz to simulate a system using as few parameters is based on the idea that the bond dimension can be truncated if a truncation in the entanglement spectrum of the reduced system still yields a sufficiently good approximation \cite{peschel1999density, perez2006matrix, klumper1993matrix}.

This leads us to the question: in our case, is there a $O(K)$ truncation of the $M^K$ space such that we still have a suitably good approximation with low error for $\langle m_{s'}\rangle$? If the $\epsilon_\alpha$ are sufficiently spread out, then the sum over $\alpha$ can be truncated to only a few terms while giving a good estimate of the $\langle m_{s'}\rangle$. To answer this question, the theory of entanglement Hamiltonians  provides a way of explicitly computing $\rho_K$ by considering the subsystem to be immersed in a thermal bath corresponding to the rest of the system. Under this formalism, we can write $\rho_K$ as a thermal ensemble under an entanglement Hamiltonian $H_K$ with eigenvalues ${\epsilon_\alpha}$:

\begin{equation}
    \rho_K=\frac{e^{-\beta H_K}}{Z};\quad Z=\mathrm{Tr}[e^{-\beta H_K}]=\sum_\alpha e^{-\epsilon_\alpha} \label{eq:CFT}
\end{equation}
where $\beta$ is the entanglement inverse temperature. In our case, $\beta$ and $H_K$ are obtainable using results from conformal field theory (CFT)  \cite{Mendes-Santos2019}, with the details discussed in Supplement \ref{supp:entanglement}.

In this framework, the trace of $e^{-\beta H_K}$ acts like a partition function at inverse temperature $\beta$. Using this result, we can determine whether the partition function can be effectively truncated to only $O(K)$ terms, analogous to the truncation of the entanglement spectrum necessary for MPS and DMRG \cite{peschel1999density, perez2006matrix, klumper1993matrix}.

The key factor is the value of $\beta$, which is fixed by the conformal field theory for the $N$ particle Hamiltonian and ground state $\ket{\Psi_{GS}}$. In the case of low effective temperature $(\beta \gg 1)$, most of the partition function $Z$ is concentrated in only a few low-lying eigenvectors $\ket{\alpha}$. At the opposite extreme of high effective temperature $(\beta\ll 1)$, the eigenvalues of $\rho_K$ will be close to uniformly distributed, requiring summing over all the eigenvectors to estimate $\braket{ m_{s'}}_{GS}$. As shown in Fig.~\ref{fig:alphacutoff}, this case happens to be in the ``low-temperature'' regime. This means that despite having $O\left(2^K\right)$ total contributions to the partition function, only $O(K)$ of them are sufficient to capture the behavior of the subsystem.


The results above give an argument why there exists a low-dimensional approximation to the exact ground state, and why it can be accessed by an ansatz that deals only with motifs. In addition, we conjecture that this process generalizes to other systems, where the effective temperature determines the truncation of the entanglement spectrum in a particular basis, which then controls how well an ansatz in that basis can effectively describe the ground state using only linearly many parameters.

Finally, we note that any substrings that are connected to each other by a symmetry of $H_K$ will have the same MEV values since $\braket{m_{s'}}=\braket{s'|e^{-\beta H_K}|s'}/Z$. Thus, we can also group motifs symmetric to each other into motif ``equivalence classes''. We will use these motif classes later in Sec 4.

\section{Statistical \& Physical Interpretations of the CNN Ansatz}
\subsection{Motivation}
The previous section demonstrates why and when it is possible for a generic variational ansatz to approximate the ground state using only linearly many parameters. We now turn our attention to the CNN's functional form and the specific nature of its approximation.
Traditionally, variational ansatzes in physics would be derived by first focusing on a desired physical property in the target solution, and then constructing a variational ansatz with the mathematical form that allows it to capture that property. The same is true of neural networks: for example, PauliNet \cite{hermann2020deep} and FermiNet \cite{spencer2021learning} enforce the fermionic Pauli Exclusion principle in their functional forms in order to solve fermionic many body systems. So what physics does the CNN ansatz's structure allow it to capture better? We have already shown that the CNN ansatz is based on the motif counts for the input string, and can be rewritten as:
\begin{equation}
\psi^\mathrm{CNN}(s)=\prod_i  \exp\left[v\sigma ( w s_{i:i+K-1}+b)\right]=\exp\left[\sum_{s'}\mathcal{C}_{s'} m_{s'}(s)\right], \label{eq:CNNmotif}
\end{equation}
where $\mathcal{C}_{s'} \equiv \ln \phi(s')$. Thus it is both an \emph{exponential} and \emph{product} ansatz.

\subsection{CNN as a Maximum Entropy (MaxEnt) Ansatz}
As mentioned before, CNNs cannot directly `see' the full $N$-string and only operate based on the information given about the $K$-motifs. The most natural ansatz then to compare the CNN to is a Maximum Entropy (MaxEnt) ansatz constrained by motif expectation values. We do so by exploiting the positive definiteness of the ground state wavefunction to define the classical probability distribution $P(s)=|\psi^{\textrm{MaxEnt}}(s)|^2$. The only information (and thus set of constraints) imposed on this MaxEnt ansatz is that the $K$-motif expectation values (MEVs) $\braket{m_{s'}}=q(s')= \sum_s P(s) m_{s'}(s)$
should match those of the ground state wavefunction.
\begin{equation}
    \braket{m_{s'}}=\braket{m_{s'}}_{GS} \Longleftrightarrow q(s')= q_{GS}(s') \quad \forall s'
\end{equation}Note that $q(s')$ is also the $K-$ marginal distribution for the joint probability $P(s)$. This results in a MaxEnt ansatz for our wavefunction \cite{Canosa1989}
\begin{equation}
    \ln\psi^{\text{MaxEnt}}(s)=\sum_{s'}\lambda_{s'} m_{s'}(s)
\end{equation} 
where each $\lambda_s'$ is a Lagrange multiplier associated with the constraint on $\braket{m_{s'}}$.
It is immediately apparent that the classical MaxEnt ansatz constrained on $K-$ marginals using motif expectation values has the same analytical form as our CNN ansatz. Both are exponential distributions, and both are agnostic to the full input $s$ other than the motif frequencies for $s'$ in $s$. This observation points out to the fact that the CNN model is more than just an empirically data-driven choice for solving quantum spin problems and in fact, has a much deeper significance than previously imagined. By definition, the MaxEnt ansatz is maximally indifferent' to everything except the constrained observables, which makes the CNN a natural choice to solve this class of problems.

As an aside, we can also consider constraints on the symmetry of the wavefunction for an arbitrary symmetry group $G$. The solution, in this case, follows the same procedure as above, but with a basis of equivalence classes, where the observables are defined as the average over equivalence classes (equations 27-31 of \cite{losada2019solutions}). This yields the MaxEnt ansatz over each equivalence class $\mathcal{E}_k$ as (see Sec. \ref{suppsec:derive_maxent}):
\begin{equation}
    \ln\psi^{\mathrm{MaxEnt}}(\mathcal{E}_k)=\sum_{s'}\lambda_{s'} \tilde m_{s'}(\mathcal{E}_k)+ \ln Z
\end{equation}
where $\ln Z$ is a renormalization term.

\subsection{CNN as an Entangled Plaquette Correlator Product State (EP-CPS) Ansatz }
The CNN ansatz as written in Eqs.~\ref{eq:CNNphi} and \ref{eq:CNNmotif} is a product of a function $\phi(s')$ with multiplicity determined by the motif counting operator. In this form, the CNN ansatz is similar to a correlator product state (CPS) ansatz as defined by a product of correlator parameters, but with extra constraints on the functional form of the parameters. The CPS ansatz has been already seen widespread use in 1D and 2D spin systems, and its wavefunction is given by

\begin{table}
\centering
\begin{tabularx}{\textwidth} { 
  | >{\centering\arraybackslash}X 
  | >{\centering\arraybackslash}X 
  | >{\centering\arraybackslash}X 
  | >{\centering\arraybackslash}X | }\hline
    \textbf{Method/Ansatz} & CNN & CPS & MaxEnt \\\hline
    \textbf{Field} & Machine learning & Physics & Statistics\\\hline
   \textbf{Coefficients $\mathcal{C}_{s'}$} & $v \sigma(ws'+b) $& $\ln \phi(s')$ & Lagrange Multipliers $\lambda_{s'}$\\\hline
    \textbf{Functional Form} & Product and Exponential & Product & Exponential \\\hline
    \textbf{Training Goal} & Minimize Energy & Any & Maximize Entropy \\\hline
    \textbf{Known Information } & Hamiltonian & Effective range of interactions & Moment Constraints\\\hline
    \textbf{Parameters} & $v$, $w$, $b$ & $\phi$ & $\lambda$\\\hline
    \textbf{Hyperparameters} & filter size $K$ & plaquette size $K$ & $K$-marginal distributions preserved\\\hline
    \textbf{Activation Function} & nonlinearity $\sigma(\cdot)$ & dependencies b/w coupling constants & dependencies b/w Lagrange multipliers  \\\hline
\end{tabularx}
\caption{Connections between CNN, CPS, and MaxEnt ansatzes. The equation unifying these three frameworks is given by $\ln \psi^{\mathrm{ansatz}}=\sum_{s'\in \mathcal{M}_{M,K} }\mathcal{C}_{s'} m_{s'}(s)$}
\label{tbl:unifying-povs}
\end{table}

\begin{equation}
    \psi^\mathrm{CPS}(s)=\prod_i\phi_{s_{i:i+K-1}}
\end{equation}
The CPS ansatz therefore maps onto the CNN ansatz when the correlator parameters $\phi_{s'}$ are set to $e^{v\sigma(ws'+b)}$. These results are summarized in the following theorem:

\begin{theorem}The CNN acts as both a restricted CPS ansatz and a MaxEnt ansatz subject to constraints on the MEVs. Furthermore, the CPS ansatz can be mapped onto a MaxEnt ansatz with  $K-$marginal moment constraints.
\end{theorem}
The unification between the CNN, CPS, and MaxEnt wavefunctions is shown in Table \ref{tbl:unifying-povs}, where each ansatz is expressed as
\begin{equation}
    \ln \psi=\sum_{s'}\mathcal{C}_{s'} m_{s'}(s). \label{eq:unify}
\end{equation}

\subsection{Numerical Evidence: CNNs Behave like Restricted CPS Approximations} \label{sec:reg_all}
\begin{figure} [t]
    \centering
    \includegraphics[width=\textwidth]{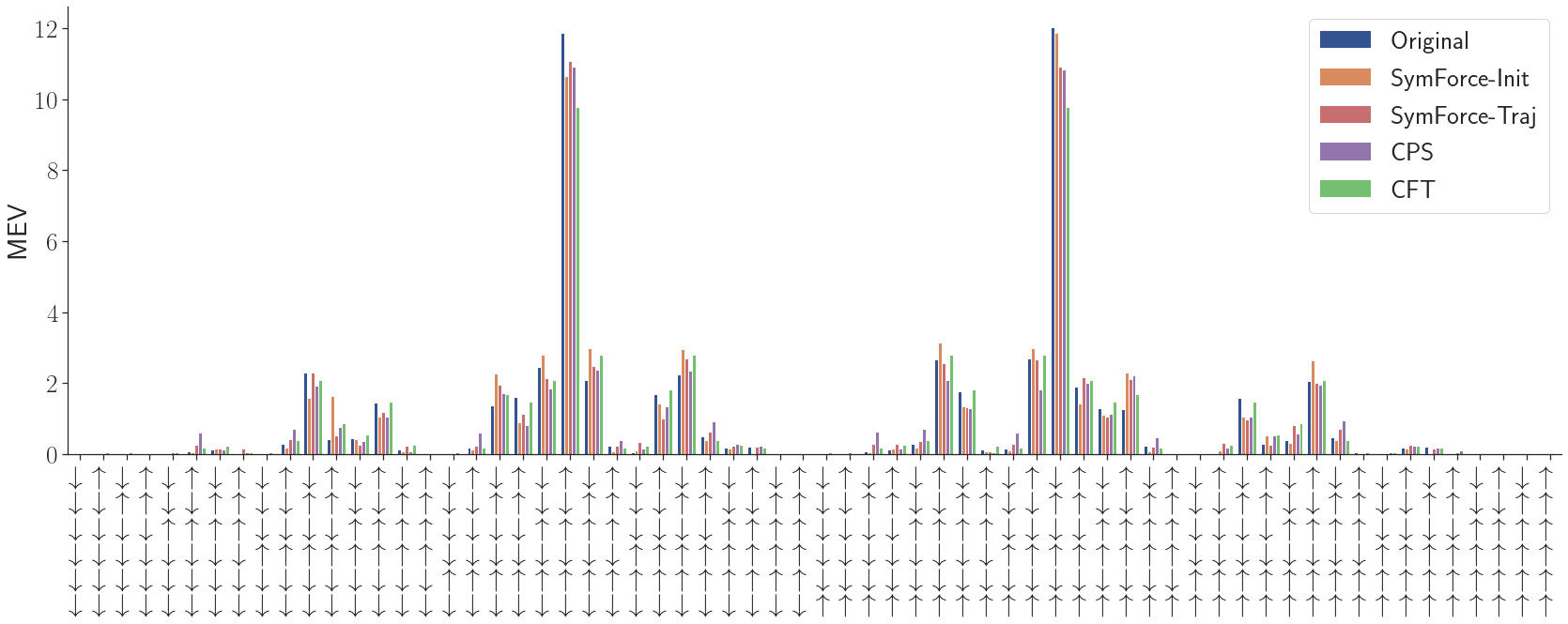}
    \caption{Motif Occurrence Expectation Values (MEVs) learned by the original algorithm, SymForce-Init, SymForce-Traj and CPS, compared to the CFT values as the ground truth ($N = 60$). We choose the best run for each algorithm in terms of the relative error $\delta_E = (\hat{E}_0 - E_0) / (E_1 - E_0) \times 100\%$. $\delta_E^{\mathrm{Original}} = 16.7\%$, $\delta_E^{\mathrm{SymForce-Init}} = -2.27\%$, $\delta_E^{\mathrm{SymForce-Traj}} = 7.37\%$ and $\delta_E^{\mathrm{CPS}} = 6.32\%$.
    }
    \label{fig:mev_all}
\end{figure}

We examine the similarities and differences between the CNN and CPS ansatzes by training CPS models and comparing the learned MEVs under the best hyperparameters for each algorithm. We use the CFT values (see Sec. \ref{sec:entanglement}) as the ground truth. We adapt the variational CNN training scheme for CPS models parameterized using Eq.~\eqref{eq:unify} and set $\mathcal{C}_{s'}$, $\forall s^\prime \in \mathcal{M}_{M, K}$ as the trainable parameters (see Sec. \ref{suppsec:hyperparameters_bar} for hyperparameters). In Fig.~\ref{fig:mev_all}, we can see that the original shallow CNN overestimates the MEVs for the Neel motifs while underestimates the MEVs for motifs that are very close to the Neel motifs (eg. $\downarrow \uparrow \uparrow \downarrow \uparrow \downarrow$, $\downarrow \uparrow \downarrow \uparrow \uparrow \downarrow$, etc.). In contrast, the MEVs learned by our symmetry-forcing algorithms have better alignment with the CPS model. All learned MEVs, however, deviate a little from the CFT values, as $\delta_E \ne 0$.

To further compare the CNN and CPS model families through their behavior under different hyperparameters, for each model, we perform a separate regression of $\delta_E$ (see Table \ref{tab:regression_all}) on (i) relative errors in MEVs $\bar{\delta}_0$ and $\bar{\delta}_1$, and (ii) kernel size $K$ (other hyperparameters are not statistically significant). We define $\delta_E = (\hat{E}_0 - E_0) / (E_1 - E_0)$, which is the error between the predicted and actual ground state energy relative to the energy gap between the first excited state and the ground state (See Sec. \ref{suppsec:reg_all} for details on $E_1$). As for $\bar{\delta}_k$ for $k \in \{0, 1\}$, it represents the relative error in MEVs ($\delta_k = (\langle m_{s_k^\prime} \rangle - \langle m_{s_k^\prime} \rangle_{GS}) / \langle m_{s_k^\prime} \rangle_{GS}$) averaged over motifs $s_k^\prime$ in the motif equivalence class (defined at the end of Sec. 3) with the $k$th highest MEVs.

We find that CNNs behave similarly to the CPS ansatz under different hyperparameters, since all four regressions have high $R^2$ around $0.5 \sim 0.6$, and the intercept and the coefficients for $\delta_1$ have the same sign and similar magnitudes. The coefficients of $K$ are larger for CNNs than that for the CPS ansatz. This further implies that a CNN behaves like a restricted CPS model, which needs a much larger $K$ to be able to model the MEVs. It is worth noting the symmetry-forcing algorithms are less sensitive to $K$, indicating that they have better robustness to hyperparameter choices.


 \begin{table}[]
    \begin{center}
        \caption{Regression results for $\delta_E$ vs. $\bar{\delta}_k$, $k \in \{0, 1\}$, without outliers ($\delta_E \geq 6$).}
        \begin{tabular}{rrrrr}
            \toprule
                    \textbf{Algorithm} &       \textbf{Original} &  \textbf{SymForce-Init} & \textbf{SymForce-Traj} &            \textbf{CPS} \\
            \midrule
     $R^2$ &                 0.627 &                 0.651 &                 0.594 &                 0.479 \\
  No. Obs. &                    27 &                    37 &                    36 &                    40 \\
 Cond. No. &                  33.0 &                  35.9 &                  42.0 &                  38.3 \\
 Intercept &  $2.852$*** $(0.458)$ &  $1.918$*** $(0.266)$ &  $2.181$*** $(0.248)$ &   $0.811$** $(0.255)$ \\
$\bar{\delta}_0$ &    $-0.655$ $(0.375)$ &     $0.466$ $(0.236)$ &    $-0.022$ $(0.307)$ & $-1.043$*** $(0.284)$ \\
$\bar{\delta}_1$ &  $-2.851$** $(0.840)$ & $-2.461$*** $(0.495)$ &    $-0.976$ $(0.567)$ & $-1.762$*** $(0.472)$ \\
       $K$ & $-0.375$*** $(0.092)$ & $-0.243$*** $(0.053)$ & $-0.294$*** $(0.048)$ &    $-0.086$ $(0.051)$ \\
            \bottomrule
        \end{tabular}
        \label{tab:regression_all}
    \end{center}
    Standard errors are reported in parentheses after the coefficients. * indicates significance at the 95\% level. ** indicates significance at the 99\% level. *** indicates significance at the 99.9\% level.
\end{table}

\section{Physical Insights from Learned CNNs}
We have demonstrated that the CNN is fundamentally an ansatz built around the motif counting operator $m_{s'}(s)$ and its expectation values. The values of this operator form the entries of the motif count matrix. We've shown that the motif  expectation values also match entries of the reduced density matrix $\rho_K$ and shown that the reduced density matrix can be well-approximated using only $O(K)$ summation terms.

We now explore a related question: what physical features of the system do the expectation values $\braket{m_{s'}}$ correlate with? And does there exist a lower-dimensional structure i.e. with fewer intrinsic degrees of freedom? To illustrate an example of this analysis, we pick two salient features of the Hamiltonian: a) $n_\mathrm{like} (m)$, the number of pairs of adjacent like spins, and b) $d_{\mathrm{Neel}}(m)$, the edit distance from each motif to the closest Neel motif. 

\begin{wraptable}{r}{8cm}
    \begin{center}
        \caption{Regression Results for MEVs vs. Physical Features of Interest. $R^2 = .786$, \# Obs. $= 64$, Cond. No. $= 42.4$.}
        \label{tab:reg_moment_phy}
        \begin{tabular}{rr}
            \toprule
            Variable & Coefficient (std.) \\
            \midrule
            Intercept & $8.95\text{*}(0.60)$ \\
            $d_{\mathrm{Neel}}$ & $-2.97\text{*}(0.35)$ \\
            $n_{\mathrm{like}}$ & $-3.60\text{*}(0.30)$ \\
            $d_{\mathrm{Neel}} \cdot n_{\mathrm{like}}$ & $1.25\text{*}(0.13)$ \\
            \bottomrule
        \end{tabular}
    \end{center}
    * indicates significance at the 99.9\% level.
\end{wraptable}

Both these metrics are rooted in physical observations of the system. The Sutherland model eigenspectrum can be thought of as one that favors having unlike pairs in basis states and penalizes having like pairs. The Neel states with the highest $\ln \psi(s)$ are the states with the least number of like pairs, whereas the ferromagnetic states with the lowest $\ln \psi(s) $ have the highest number of like pairs. Similarly, the Sutherland Hamiltonian can be thought of as generating swaps of adjacent spins at position $i$, and all basis states of the Sutherland system can be generated using iterative local swaps from the Neel states. 

We focus on the $N=60$ system and examine the CNN with a kernel size $K = 6$ with the best training hyperparameters, algorithm, and random seed, which has $\delta_E = -2.27\%$ (defined as in Sec. \ref{sec:reg_all}). In Table \ref{tab:reg_moment_phy}, we show our best regression model. With a high $R^2=0.786$, the model shows that both a larger edit distance from the Neel state and having more like pairs lead to lower MEVs. This effect saturates since the coefficient for the interaction term is positive but relatively small. This regression analysis reveals that MEVs, which are crucial to the CNN, have a much simpler dependence on the physics of the system than anticipated and that an accurate low-dimensional approximation of the CNN exists.

\section{Discussion }
Our goals in this paper are twofold: determining how the CNN circumvents the ``Curse of Dimensionality'' and understanding the nature of the CNN's low dimensional approximation. Our results in this paper offer several clues into answering these questions. Firstly, we show how the symmetries of the target wavefunction constrain the CNN during training (and offer a new algorithm that optimizes training by explicitly forcing these constraints). Our symmetry enforced algorithms give similar performance to a mult-layer deep convolutional neural network. In fact we show that adding layers to the original network doesn't necessarily always give training gains, but the symmetry enforced algorithms do.  We then use several theoretical tools to establish that the key to the CNN's low dimensional ansatz lies in understanding the convolutional operator through the motifs and the motif expectation values. This reinterpretation of the CNN enables us to understand the power and limits of the CNN to approximate a system in $O(K)$ parameters from the lens of entanglement spectrum theory, and allows us to interpet the CNN's functional form in terms of a MaxEnt ansatz as well as connect it to a CPS ansatz. We then conduct regression analyses to demonstrate that the CNN cuts corners by focusing primarily on a few top MEVs, and how this can be interpreted using the connection to MaxEnt as well as a truncation of the reduced density matrix and entanglement spectrum of the $K-$ subsystem. And finally, we use another regression analysis to show how the MEVs strongly depend on a few physical insights about the Hamiltonian, suggesting further lower dimensional structure that the CNN is focusing on for its approximation. While we conducted our analysis for the Sutherland Hamiltonian, many of the concepts we introduced may be applied to investigate how CNNs approximate other spin Hamiltonians.

\section*{Acknowledgments}
 H.P. acknowledges support from NSF (Grant No. PHY-2207283) and the Welch Foundation (Grant No. C-1669).
 A.B.P. and Y.J acknowledge support from NSF (Grant No. DBI-1707400) and NIH (Grant No. P42ES027725)
 F.A. and A.B.P. were supported by the Intelligence Advanced Research Projects Activity (IARPA) via Department of Interior/Interior Business Center (DoI/IBC) contract no. D16PC00003. The US Government is authorized to reproduce and distribute reprints for governmental purposes notwithstanding any copyright annotation thereon. This work is also supported by the Lifelong Learning Machines (L2M) Program of the Defense Advanced Research Projects Agency (DARPA) via contract number HR0011-18-2-0025 and R01 EY026927 to AT and by NSF NeuroNex grant 1707400.

\newpage
\appendix

\section{Representation, Dynamics and Inductive Bias of CNN}
In this section, we provide detailed proofs or derivations of statements mentioned in the main text. 
\subsection{Derivation of Motif Count Vectors} \label{supp:lnpsi-CNN-string}
Eq.~\eqref{eq:CNNphi} is derived from Eq.~\eqref{eq:CNN_arch} as following:
\begin{align}
    \psi^\mathrm{CNN}(s) =& \exp \Big[ v \sum_{i=1}^N \sigma ( w s_{i:i+K-1}+b)\Big], \nonumber \\
    =& \exp \Big[ v \sum_{i=1}^N\sigma ( w s_{i:i+K-1}+b)\Big], \nonumber \\
    =& \exp \Big[ v \sum_{i=1}^N\sum_{s^\prime \in \mathcal{M}_{M, K}} \delta_{s_{i:i+K-1},s^\prime} \cdot \sigma ( w s^\prime+b) \Big], \nonumber \\
    =&  \exp \Big[ v \sum_{s^\prime \in \mathcal{M}_{M, K}} \sigma ( w s^\prime+b) \cdot  \sum_{i=1}^N  \delta_{s_{i:i+K-1},s^\prime} \Big], \nonumber \\
    =&  \exp \Big[ v \sum_{s^\prime \in \mathcal{M}_{M, K}} \sigma ( w s^\prime+b) \cdot m_{s^\prime}(s) \Big], \nonumber \\
    =& \prod_{s^\prime \in \mathcal{M}_{M, K}} \exp \Big[ v  \sigma ( w s^\prime+b) \cdot m_{s^\prime}(s) \Big], \forall s \in \mathcal{S}_{N, M}, \label{eq:lnpsi-CNN-motif-supp}
\end{align}
where $m_{s^\prime}(s) \in \mathbb{N}$ is the number of occurrence of motif $s^\prime$ in state $s$.

\subsection{The Motif Count Matrix and Critical Kernel Size} \label{appendix:critical-kernel-size}
In this section, we justify the need for a sufficiently large kernel size $K$ in a CNN. Let $a(s^\prime) \equiv \sigma ( w s^\prime+b)$ denote the motif activation. Then, Eq.~\eqref{eq:lnpsi-CNN-motif-supp} can be written as
\begin{align}
    \ln \psi^\mathrm{CNN}(s) = v \sum_{s^\prime \in \mathcal{M}_{M, K}} a(s^\prime)m_{s^\prime}(s) \equiv vc(s)^Ta, \label{eq:motif_act}
\end{align}
where $a$ is the concatenation of $a(s^\prime)$, for all $s^\prime \in \mathcal{M}_{M, K}$, and $c(s)$ is the concatenation of $m_{s^\prime}(s)$, for all $s^\prime \in \mathcal{M}_{M, K}$. Then, if we concatenate $\ln \psi^\mathrm{CNN}(s)$ for all $s \in \mathcal{S}$, we can write Eq.~\eqref{eq:motif_act} in a vectorized form by defining $C \equiv [\cdots \mid c(s) \mid \cdots]$ and $\ln \Psi \equiv [\cdots \mid \ln \psi^\mathrm{CNN}(s) \mid \cdots]^T$. Then, we have 
\begin{align}
    \ln \Psi^\mathrm{CNN} = vC^Ta \equiv C^Ta^\dagger, \label{eq:motif-mat-ls}
\end{align}
Since the prediction $\ln \Psi$ is obtained from a matrix multiplication involving the motif count matrix $C$, with $v$ and $a$ being relatively free parameters, the number of distinct values in $\ln \Psi$ is upper bounded by $\mathrm{rank}(C_{N,K})$ (here we fix $M$). Let $|\mathcal{E}_N|$ denote the number of equivalence classes in $\mathcal{S}$. Then, a CNN capable of expressing all equivalence classes exactly should have a kernel size of at least $K^*$ such that $\mathrm{rank}(C_{N,K^*}) \geq |\mathcal{E}_N|$. In Fig.~\ref{fig:classcutoff} (top), we show how $K^*$ grows along with $N$ when $M=2$. We observe a superlinear rate, indicating the problem becomes much more complicated with larger $N$ if we seek an exact solution.

\textit{Proof.} Assume $a^\dagger$ is not constrained. Then, Eq. (\ref{eq:motif-mat-ls}) is a linear system where there are $|M^K|$ unknown variables in $a^\dagger$. Let $\Tilde{C} = [C^T \mid \ln \Psi]^T$ be the augmented matrix of the linear system. Let $r=\mathrm{rank}(C)$ and $\Tilde{r}=\mathrm{rank}(\Tilde{C})$. Let $\#\mathrm{solns}$ denote the number of exact solutions. Then, according to linear algebra theory \cite{strang2016introduction}:
\begin{align}
    \#\mathrm{solns} =\left\{
    \begin{aligned}
        0 &, \quad r = \tilde{r} < |\mathcal{E}| \\
        \infty &, \quad r = \tilde{r} \geq |\mathcal{E}| \\
        0 &, \quad r < \tilde{r},
    \end{aligned}
    \right. \label{eq:ac} 
\end{align}

Hence, $r \geq |\mathcal{E}|$ is required if an exact solution is desired. Since for a given system with $N$ sites and $M = 2$ local spin states, $r$ is a function of the kernel size $K$, we simply define $K^* = \mathrm{argmin}_{K} \text{ } r(K)$, $s.t.$ $r(K) \geq |\mathcal{E}|$. \qed

In fact, we can easily show by construction that for any value of $M$, $K^*\geq\lfloor\frac{N}{3}\rfloor$. For $K<\frac{N}{3}$, we can construct a string $A - x - A - y - A - z$, where $A$ has length $K-1$ and $x$, $y$, and $z$ each have length at least 1, and have a combined length of 3, 4, or 5, depending on the value of N (mod 3). By exchanging $x$ and $y$, we obtain a second string $A - y - A - x - A - z$, which is not related to the first string by any symmetry. However, these two strings have exactly the same motifs of size $K$. Therefore, $K^*>K\geq\frac{N-2}{3}=\lfloor\frac{N}{3}\rfloor$.

\subsection{The Need for Nonlinear Activation Function in CNN} \label{ap:sec:failure}
For this particular problem, we found that a non-linear activation function is needed to prevent the CNN from producing constant outputs. We have the following Theorem.

\begin{theorem}
    If a one-layer one-filter CNN uses a linear activation function, the CNN output $ln \psi(s)$ is the same for every input state $s$. Specifically, we have $ln \psi(s) = v\langle a, c(s) \rangle = Nv\mathrm{grandsum}(\tilde{w})/M$, where $\tilde{w} = w + b / K$.
\end{theorem}
\textit{Proof.}
In the case where we use linear activation $\sigma(x) = x$, we have 
\begin{align}
    \langle a, c(s) \rangle =& \sum_{s^\prime \in \mathcal{M}} m_{s^\prime} (s) a(s^\prime) = \sum_{s^\prime \in \mathcal{M}} m_{s^\prime} (s) \sigma \big(\langle w, s^\prime \rangle + b\big) = \sum_{s^\prime \in \mathcal{M}} m_{s^\prime} (s) \big(\langle w, s^\prime \rangle + b\big) \nonumber \\
    =& \sum_{s^\prime \in \mathcal{M}} m_{s^\prime} (s) \langle \tilde{w}, s^\prime \rangle = \Big\langle \tilde{w}, \sum_{s^\prime \in \mathcal{M}} m_{s^\prime} (s) s^\prime \Big\rangle, \label{supp:eq:linear_act}
\end{align}
In order the proceed, we break down each motif $s^\prime \in \mathbb{R}^{M \times K}$ into columns vectors $\big( s_1^\prime, s_2^\prime, \cdots s_K^\prime \big)$. Then, we define the set of spins $\mathcal{A} = \{\uparrow, \downarrow, \cdots\}$ where $|\mathcal{A}| = M$. For each spin $\mathcal{A}_i \in \mathcal{A}$, we define its one-hot representation as $x_{\mathcal{A}_i}$. For example, in the case of $M=2$, we have $\mathcal{A} = \{\uparrow, \downarrow\}$, $x_{\mathcal{A}_1} = x_{\uparrow} = (1, 0)^T$ and $x_{\mathcal{A}_2} = x_{\downarrow} = (0, 1)^T$.

Then, the summation term in Eq.~\eqref{supp:eq:linear_act} becomes
\begin{align}
    \sum_{s^\prime \in \mathcal{M}} m_{s^\prime} (s) s^\prime =& \Big(\sum_{s^\prime \in \mathcal{M}} m_{s^\prime} (s) s_1^\prime, \sum_{s^\prime \in \mathcal{M}} m_{s^\prime} (s) s_2^\prime, \cdots,  \sum_{s^\prime \in \mathcal{M}} m_{s^\prime} (s) s_K^\prime) \nonumber \\
    =& \Big(\sum_{s^\prime \in \mathcal{M}} m_{s^\prime} (s) \sum_{i=1}^M \delta(s_1^\prime = x_{\mathcal{A}_i}) x_{\mathcal{A}_i}, \sum_{s^\prime \in \mathcal{M}} m_{s^\prime} (s) \sum_{i=1}^M \delta(s_2^\prime = x_{\mathcal{A}_i}) x_{\mathcal{A}_i}, \nonumber \\
    &\cdots,  \sum_{s^\prime \in \mathcal{M}} m_{s^\prime} (s) \sum_{i=1}^M \delta(s_K^\prime = x_{\mathcal{A}_i}) x_{\mathcal{A}_i} \Big) \nonumber \\
    =& \Big(\sum_{i=1}^M  x_{\mathcal{A}_i} \sum_{j=1}^N \delta \big((T^i s)_1 = x_{\mathcal{A}_j}\big), \sum_{i=1}^M  x_{\mathcal{A}_i} \sum_{j=1}^N \delta \big((T^i s)_2 = x_{\mathcal{A}_j}\big), \nonumber \\
    &\cdots, \sum_{i=1}^M  x_{\mathcal{A}_i} \sum_{j=1}^N \delta \big((T^i s)_K = x_{\mathcal{A}_j}\big) \Big), \label{supp:eq:linear_act_2}
\end{align}
where $T^i$ is the translation operator of $i$ units. The last equal sign holds because the $\delta$ function is just trying to match each individual column of the full string $s$ to each one-hot representation of the spins. Then, we know that there are exactly $N/M$ spins for each spin in the state. Therefore, Eq.~\eqref{supp:eq:linear_act_2} becomes
\begin{align}
    &\Big(\sum_{i=1}^M  x_{\mathcal{A}_i} \sum_{j=1}^N \delta \big((T^i s)_1 = x_{\mathcal{A}_j}\big), \sum_{i=1}^M  x_{\mathcal{A}_i} \sum_{j=1}^N \delta \big((T^i s)_2 = x_{\mathcal{A}_j}\big), \cdots, \sum_{i=1}^M  x_{\mathcal{A}_i} \sum_{j=1}^N \delta \big((T^i s)_K = x_{\mathcal{A}_j}\big) \Big) \nonumber \\
    =& \Big(\frac{N}{M}\sum_{i=1}^M  x_{\mathcal{A}_i}, \frac{N}{M}\sum_{i=1}^M  x_{\mathcal{A}_i}, \cdots, \frac{N}{M}\sum_{i=1}^M  x_{\mathcal{A}_i} \Big) = \frac{N}{M} \mathbf{1}_{M\times K},
\end{align}
where the last equal sign holds simply because the sum of all one-hot representations is an all-one vector. And immediately, for Eq.~\eqref{supp:eq:linear_act}, we have $\langle a, c(s) \rangle = \langle \tilde{w}, \frac{N}{M} \mathbf{1}_{M\times K} \rangle = N \mathrm{grandsum}(\tilde{w}) / M$.

Note that since we are using the ReLU activation function in the actual training, there are also cases where  ReLU is effectively linear: 1) all neurons are firing, i.e., the preactivations are all positive, and 2) all neurons are muted (not firing), i.e., the preactivations are all negative and the activations are all 0s. Hence, this proves that when all the neurons in the CNN are firing or muted (not firing), the CNN output is a constant for any input state. \qed

\subsection{The Grand Sum Condition} \label{supp:sec:gs}
\paragraph{Group theory interpretation: Equivalence to Convolution.} We find that the expression in Eq.~\eqref{eq:CNN_arch} can be rewritten in terms of convolution w.r.t. the cyclic group:
\begin{eqnarray}
\frac{\ln \psi(s)}{v} &=&
\sum_{i=1}^N \sigma \Big(\langle w, s_{i:i+K-1} \rangle + b \Big) \nonumber \\
&=& \sum_{i=1}^{N}\sigma(\langle T^{i}w, s \rangle+b) \nonumber \\ 
&=& j^{T}\sigma(w * s+b), \label{eq:group_trans}
\end{eqnarray}
where we slightly abuse $w$ to denote a kernel of size $N \times M$ instead of $K \times M$, starting from the second line in Eq.~\eqref{eq:group_trans} and $supp(w)=[1,\cdots,K]$. $T$ is the generator of cyclic shifts of the group $\mathcal{C}$, the cyclic group, and $j$ is the all-ones vector. We define $\sigma:\mathbb{R} \to \mathbb{R}$ as the non-linearity function. In the third line of Eq.~\eqref{eq:group_trans} we also slightly abuse it as function from $\mathbb{R}^{N \times M}$ to $\mathbb{R}^{N \times M}$, with element-wise nonlinearity. We use this alternative $w \in \mathbb{R}^{N \times M}$ in the proof of the grand sum condition.

\paragraph{\textit{Proof} for Thm.~\ref{thm:gs}.} In the following derivation, we use one-hot representation of the strings, i.e. a string $s \in \{0, 1\}^{N \times 2}$ (consider the case where $M = 2$), with each row having exactly one 1 and one 0. We also have the kernel weight $w \in \mathbb{R}^{N \times 2}$ and bias $b \in \mathbb{R}$. Let $s_i$ and $w_i$ denote the $i$th row of $s$ and $w$. Then, $\langle w, s \rangle$ is defined as $\mathbf{tr}(w^Ts)$. We define $\mathrm{grandsum}$ of a vector or a matrix to be the sum of all its elements.
Let $\tilde{w} = w + b / N$. We can rewrite Eq. (\ref{eq:group_trans}) as
\begin{align}
    \frac{\ln \psi(s)}{v} 
    =& \sum_{i=1}^N \sigma \big( \langle T^i w, s \rangle + b \big) \nonumber \\
    =& \sum_{i=1}^N \sigma \bigg( \sum_{j=1}^N \langle w_{j-i}, s_j \rangle + b \bigg) \nonumber \\
    =& \sum_{i=1}^N \sigma \bigg( \sum_{j=1}^N \langle w_{j-i} + b / N, s_j \rangle \bigg) \nonumber \\
    =& \sum_{i=1}^N \sigma \bigg( \sum_{j=1}^N \langle \tilde{w}_{j-i}, s_j \rangle \bigg) \nonumber \\
    =& \sum_{i=1}^N \bigg( \sum_{j=1}^N \langle \tilde{w}_{j-i}, s_j \rangle - \beta_i \langle T^i \tilde{w}, s \rangle \bigg) \nonumber \\
    =& \sum_{j=1}^N \bigg \langle \sum_{i=1}^N \tilde{w}_{j-i}, s_j \bigg \rangle - \sum_{i=1}^N \beta_i \langle T^i \tilde{w}, s \rangle \nonumber \\
    (a)=& \frac{N}{2} \mathrm{grandsum}(\tilde{w}) - \sum_{i=1}^N \beta_i \langle T^i \tilde{w}, s \rangle, \label{eq:s}
\end{align}
where $\beta_i = 1$, if $\sum_{j=1}^N \langle \tilde{w}_{j-i}, s_j \rangle = \langle T^i \tilde{w}, s \rangle < 0$; otherwise $\beta_i = 0$. (a) is because 1) by circularly shifting $\tilde{w}$ and applying it to $s$, each site of $s$ sees each row of $\tilde{w}$ exactly once, 2) there are equal number of each spins in the string. Hence, the first term equals half of the sum of all weights in $\tilde{w}$ times $N$.

Similarly, for the reflected string of $Rs$, where $R$ is just a permutation matrix that flip $s$ along the rows, we have
\begin{align}
    \frac{\ln \psi(Rs)}{v}
    =& \frac{N}{2} \mathrm{grandsum}(\tilde{w}) - \sum_{i=1}^N \beta_i^\prime \langle T^i \tilde{w}, Rs \rangle, \label{eq:Rs}
\end{align}
where $\beta_i^\prime = 1$, if $\sum_{j=1}^N \langle \tilde{w}_{N+1+i-j}, s_j \rangle = \langle T^i \tilde{w}, Rs \rangle < 0$; otherwise $\beta_i^\prime = 0$. We can also write the expression of $\ln \psi(Ls) / v$ as
\begin{align}
    \frac{\ln \psi(Ls)}{v}
    =& \frac{N}{2} \mathrm{grandsum}(\tilde{w}) - \sum_{i=1}^N \beta_i^{\prime\prime} \langle T^i \tilde{w}, Ls \rangle, \label{eq:Ls}
\end{align}
where $L$ switches the labels of $s$ and is just a permutation matrix that flips $s$ along the columns. Here $\beta_i^{\prime\prime} = 1$, if $\langle T^i \tilde{w}, Ls \rangle = \mathrm{grandsum}(\tilde{w}) - \langle T^{i} \tilde{w}, s \rangle < 0$. When $\mathrm{grandsum}(\tilde{w}) = 0$, we have
\begin{align}
    \langle T^i \tilde{w}, Ls \rangle = - \langle T^{i} \tilde{w}, s \rangle, \label{eq:gs_Ls}
\end{align}
Thus, if $\langle T^{i} \tilde{w}, s \rangle > 0$,  then  $\beta_i^{\prime\prime} = 1$, i.e. $\beta_i^{\prime\prime} = 1 - \beta$. Then, when the parameters of a shallow ReLU CNN with 1 filter satisfies $\mathrm{grandsum}(\tilde{w})$ = $\mathrm{grandsum}(w) + 2b$ = 0, it learns the \textbf{label-switching} symmetry. To see this, take Eq. (\ref{eq:Ls}) - (\ref{eq:s}). We have
\begin{align}
    \frac{\ln \psi(Ls) - \ln \psi(s)}{v}
    =& \frac{N}{2} \mathrm{grandsum}(\tilde{w}) - \sum_{i=1}^N \beta_i^{\prime\prime} \langle T^i \tilde{w}, Ls \rangle \nonumber \\ 
    &- \left(\frac{N}{2} \mathrm{grandsum}(\tilde{w}) - \sum_{i=1}^N \beta_i \langle T^i \tilde{w}, s \rangle \right) \nonumber \\ 
    =& -\sum_{i=1}^N \beta_i^{\prime\prime} \langle T^i \tilde{w}, Ls \rangle + \sum_{i=1}^N \beta_i \langle T^i \tilde{w}, s \rangle \nonumber \\
    =& \sum_{i=1}^N (1 - \beta_{i}) \langle T^{i} \tilde{w}, s \rangle + \sum_{i=1}^N \beta_i \langle T^i \tilde{w}, s \rangle \nonumber \\
    =& \sum_{i=1}^N \langle T^i \tilde{w}, s \rangle \nonumber \\
    =& \sum_{i=1}^N \sum_{j=1}^N \langle \tilde{w}_{j-i}, s_j \rangle \nonumber \\
    =& \frac{N}{2} \mathrm{grandsum}(\tilde{w}),
\end{align}
Then, clearly $\left(\ln \psi(Ls) - \ln \psi(s) \right) / v = 0$ if $\mathrm{grandsum}(\tilde{w}) = 0$.

Next, for the reflection symmetry, we make the following assumption:
\begin{assumption}
    $\forall s$, $\exists k \in \mathbb{Z}$, $s.t.$ $T^k Ls = Rs$, where $L$ is the label switching operator. \label{as:exists_k}
\end{assumption}
Then, take Eq. (\ref{eq:Rs}) - (\ref{eq:s}), we have
\begin{align}
    \frac{\ln \psi(Rs) - \ln \psi(s)}{v}
    =& \frac{N}{2} \mathrm{grandsum}(\tilde{w}) - \sum_{i=1}^N \beta_i^\prime \langle T^i \tilde{w}, T^k Ls \rangle \nonumber \\ 
    &- \left(\frac{N}{2} \mathrm{grandsum}(\tilde{w}) - \sum_{i=1}^N \beta_i \langle T^i \tilde{w}, s \rangle \right) \nonumber \\ 
    =& \sum_{i=1}^N \beta_i^\prime \langle T^{i - k} \tilde{w}, Ls \rangle + \sum_{i=1}^N \beta_i \langle T^i \tilde{w}, s \rangle \nonumber \\
    =& \sum_{i=1}^N (1 - \beta_{i-k}) \langle T^{i - k} \tilde{w}, s \rangle + \sum_{i=1}^N \beta_i \langle T^i \tilde{w}, s \rangle \nonumber \\
    =& \sum_{i=1}^N (1 - \beta_i) \langle T^i \tilde{w}, s \rangle + \sum_{i=1}^N \beta_i \langle T^i \tilde{w}, s \rangle \nonumber \\
    =& \sum_{i=1}^N \langle T^i \tilde{w}, s \rangle \nonumber \\
    =& \frac{N}{2} \mathrm{grandsum}(\tilde{w}),
\end{align}
Then, clearly $\left(\ln \psi(Rs) - \ln \psi(s) \right) / v = 0$ if $\mathrm{grandsum}(\tilde{w}) = 0$. \qed

\subsection{Algorithms For Improving Training using the Grand Sum Condition} \label{supp:sec:alg}
Our symmetry-forcing modification can be applied to any parameter update method $\mathrm{update}(\theta)$ used during training. This modification requires $M=2$.

\begin{algorithm}[H]
\SetAlgoLined
 Let $K$ be the kernel size of $w$\;
 Calculate current value $c = \mathrm{grandsum}(w) + 2b$\;
 Project $w \leftarrow w - c / (2K)$, $b \leftarrow b$\;
 \textbf{Return} $w$, $b$
 \caption{\texttt{Project}$(w ,b)$ - Projecting CNN parameters for a 0 grand sum}
 \label{alg:proj}
\end{algorithm}

\begin{algorithm}[H]
\SetAlgoLined
 Initialize CNN parameters $\theta = (w, b, v)$ using any initializing scheme\;
 $w, b \leftarrow \texttt{Project}(w,b)$\;
 \While{\texttt{iter} $\leq$ \texttt{max\_iter}}{
  $\mathrm{update}(\theta)$\;
 }
 \textbf{Return} $\theta$
 \caption{SymForce-Init - Training CNN with the grand sum initialized as 0}
 \label{alg:CNN-init}
\end{algorithm}

\begin{algorithm}[H]
\SetAlgoLined
 Initialize CNN parameters $\theta = (w, b, v)$ using any initializing scheme\;
 $w, b \leftarrow \texttt{Project}(w,b)$\;
 \While{\texttt{iter} $\leq$ \texttt{max\_iter}}{
  $\mathrm{update}(\theta)$\;
  $w, b \leftarrow \texttt{Project}(w,b)$\;
 }
 \textbf{Return} $\theta$
 \caption{SymForce-Traj - Training CNN with the grand sum forced to be 0 at every iteration}
 \label{alg:CNN-force}
\end{algorithm}

\subsection{Learning Dynamics of CNN} \label{suppsec:learning_dynamics}
The sketch of the proof is the following: Suppose the CNN has parameters $\theta=(v,w,b)$. The updates of each $\psi(s;\theta)$ can be written as $\partial \mathcal{L} / \partial \psi(s)$, where $\mathcal{L}$ is the loss, times the neural tangent kernel (NTK) \cite{jacot2018neural}, which is a matrix whose each entry is $\partial_\theta^T\psi(s)\partial_\theta\psi(s^\prime)$. When $\psi(s) = \psi(gs)$, the loss gradient part is apparently invariant, while the NTK part becomes $\partial_\theta^T\psi(gs)\partial_\theta\psi(gs^\prime)$. After we plug in Eq.~\eqref{eq:CNN_arch}, we can show that for either $v$, $w$, or $b$, the NTK part is also invariant.
\begin{theorem}[Invariant dynamics]
If the grand sum condition is satisfied then 
\begin{equation*}
\partial_{t}\ln \psi_{\theta(t)}(s)=\partial_{t}\ln \psi_{\theta(t)}(gs),\;\;\;\forall\;g\in \{T, L\},
\end{equation*}
where $\theta(t)=(v(t),w(t),b(t))$ denotes the CNN parameters, $T$ denotes the translation and $L$ denotes the relabeling transformation.
\end{theorem}

\noindent
\textit{Proof.} The gradient flow of $\psi_{\theta(t)}$ can be written 
in the NTK formalism as:

\begin{align*}
\dot{\Psi}_{\theta(t)} = \partial_{\theta}^T \Psi \dot{ \theta} = -\partial_{\theta}^T \Psi \partial_{\theta}^T \mathcal{L} = -\Big(\partial_{\theta}^T \Psi \partial_{\theta} \Psi\Big) \partial_{\Psi}^T \mathcal{L} = -K\frac{2(||\Psi||_2^2 I - \Psi \Psi^T) H \Psi}{||\Psi||_2^4},
\end{align*}
where $K = \partial_{\theta}^T \Psi \partial_{\theta} \Psi$ is the NTK whose each entry is $\partial_\theta^T\psi(s) \partial_\theta\psi(s^\prime)$, $\forall s, s^\prime \in \mathcal{S}$. It is straightforward to see that $\partial_{\psi}^T \mathcal{L}$ is invariant w.r.t. transformations $g$ since $\psi$ is invariant under the grand sum condition. It remains to demonstrate that the kernel is invariant (i.e., whether $\partial_\theta^T\psi(s)\partial_\theta\psi(s^\prime) = \partial_\theta^T\psi(gs)\partial_\theta\psi(gs^\prime)$).
Let $\tilde{w} = w + b / N$. From Eq.~\eqref{eq:group_trans}, we have $\psi(s) = \exp\Big[v\sum_{i=1}^N \sigma(\langle T^iw, s \rangle + b)\Big] = \exp\Big[v\sum_{i=1}^N \sigma(\langle T^i\tilde{w}, s \rangle)\Big]$. Then, for the 3 different sets of variables in $\theta$:
\begin{enumerate}
    \item For $v$, we have $\partial_v\psi(s) = \psi(s) \sum_{i=1}^N \sigma(\langle T^i\tilde{w}, s \rangle) = \psi(s) \ln \psi(s) / v$. Then, since $\psi(s)$ is invariant to $g$, $\partial_v\psi(s)$ is also invariant to $g$. Hence, $\forall s, g$, we have $\partial_v^T\psi(s)\partial_v\psi(s^\prime) = \partial_v^T\psi(gs)\partial_v\psi(gs^\prime)$.
    \item For $w$, we have
    \begin{align*}
        \partial_w\psi(s) = \psi(s) v \sum_{i=1}^N \sigma^\prime(\langle \tilde{w}, T^{N-i}s \rangle) T^{N-i}s.
    \end{align*}
    Then, 
    \begin{align*}
        \partial_w^T\psi(s)\partial_w\psi(s^\prime) = \psi(s)\psi(s^\prime) v^2 \sum_{i=1}^N \sigma^\prime(\langle \tilde{w}, T^{N-i}s \rangle) \cdot \sum_{i=1}^N \sigma^\prime(\langle \tilde{w}, T^{N-i}s^\prime \rangle) s^Ts^\prime.
    \end{align*}
    Consider when applying $g$, we have
    \begin{align}
        \partial_w^T\psi(gs)\partial_w\psi(gs^\prime) = \psi(gs)\psi(gs^\prime) v^2 \sum_{i=1}^N \sigma^\prime(\langle \tilde{w}, T^{N-i}gs \rangle) \cdot \sum_{i=1}^N \sigma^\prime(\langle \tilde{w}, T^{N-i}gs^\prime \rangle) s^T(g^Tg)s^\prime. \label{eq:K_w}
    \end{align}
    It is straightforward that $\psi(gs)\psi(gs^\prime) = \psi(s)\psi(s^\prime)$ since $\psi$ is invariant to $g$. Also, $g^Tg = I$, since $g$ is unitary. To see this, consider $s \in \mathbb{R}^{2N}$ as the vectorized one-hot representation of a state when $M=2$. Then, we have
    \begin{align*}
        T = \begin{bmatrix} 
            0 & 1 & 0 & \cdots & 0 & 0 & 0 & 0 & \cdots & 0 \\
            0 & 0 & 1 & \cdots & 0 & 0 & 0 & 0 & \cdots & 0 \\
            \cdots & \cdots & \cdots & \cdots & \cdots & \cdots & \cdots & \cdots & \cdots & \cdots \\
            0 & 0 & 0 & \cdots & 1 & 0 & 0 & 0 & \cdots & 0 \\
            1 & 0 & 0 & \cdots & 0 & 0 & 0 & 0 & \cdots & 0 \\
            0 & 0 & 0 & \cdots & 0 & 0 & 0 & 0 & \cdots & 0 \\
            \cdots & \cdots & \cdots & \cdots & \cdots & \cdots & \cdots & \cdots & \cdots & \cdots \\
            0 & 0 & 0 & \cdots & 0 & 0 & 0 & 0 & \cdots & 0 \\
            0 & 0 & 0 & \cdots & 0 & 0 & 1 & 0 & \cdots & 0 \\
            0 & 0 & 0 & \cdots & 0 & 0 & 0 & 1 & \cdots & 0 \\
            \cdots & \cdots & \cdots & \cdots & \cdots & \cdots & \cdots & \cdots & \cdots & \cdots \\
            0 & 0 & 0 & \cdots & 0 & 0 & 0 & 0 & \cdots & 1 \\
            0 & 0 & 0 & \cdots & 0 & 1 & 0 & 0 & \cdots & 0 \\
	    \end{bmatrix}
	    = \begin{bmatrix} 
            T_1 & O \\
            O & T_1 \\
	    \end{bmatrix},
    \end{align*}
    where $T_1 \in \mathbb{R}^{N \times N}$ is a matrix that translates a vector by 1 unit and $O \in \mathbb{R}^{N \times N}$ is an all-zero matrix.
    Also, $L = \begin{bmatrix} 
            O & I \\
            I & O \\
	    \end{bmatrix}$,
    Then, it is trivial to check if $T$ and $L$ are unitary. Therefore, $g^Tg = I$.
    
    It remains to see if 
    \[
\sum_{i=1}^N \sigma^\prime(\langle \tilde{w}, T^{N-i}s \rangle) \cdot \sum_{i=1}^N \sigma^\prime(\langle \tilde{w}, T^{N-i}s^\prime \rangle) = \sum_{i=1}^N \sigma^\prime(\langle \tilde{w}, T^{N-i}gs \rangle) \cdot \sum_{i=1}^N \sigma^\prime(\langle \tilde{w}, T^{N-i}gs^\prime \rangle).
    \]
    For $T$, this holds simply because the summation is taken over all possible translations. For $L$, recall from Eq.~\eqref{eq:gs_Ls}, when the grand sum condition holds, we have $\langle T^i \tilde{w}, Ls \rangle = -\langle T^i \tilde{w}, s \rangle$. Hence, $\sigma^\prime(\langle T^i \tilde{w}, Ls \rangle) = 1 - \sigma^\prime(\langle T^i \tilde{w}, s \rangle)$ because the derivative of the ReLU function is 1 if the input is positive; otherwise 0. Then, by using this property, we have
    \begin{align}
        &\sum_{i=1}^N \sigma^\prime(\langle \tilde{w}, T^{N-i}Ls \rangle) \cdot \sum_{i=1}^N \sigma^\prime(\langle \tilde{w}, T^{N-i}Ls^\prime \rangle) \nonumber \\
        =& \sum_{i=1}^N \sigma^\prime(\langle T^{i}\tilde{w}, Ls \rangle) \cdot \sum_{i=1}^N \sigma^\prime(\langle T^{i}\tilde{w}, Ls^\prime \rangle) \nonumber \\
        =& \sum_{i=1}^N \Big[1 - \sigma^\prime(\langle T^{i}\tilde{w}, s \rangle)\Big] \cdot \sum_{i=1}^N \Big[1 -  \sigma^\prime(\langle T^{i}\tilde{w}, s^\prime \rangle)\Big] \nonumber \\
        =& \Big[N - \sum_{i=1}^N\sigma^\prime(\langle T^{i}\tilde{w}, s \rangle)\Big] \cdot \Big[N -  \sum_{i=1}^N \sigma^\prime(\langle T^{i}\tilde{w}, s^\prime \rangle)\Big] \nonumber \\
        =& N^2 - N \sum_{i=1}^N \Big[\sigma^\prime(\langle T^{i}\tilde{w}, s \rangle) + \sigma^\prime(\langle T^{i}\tilde{w}, s^\prime \rangle)\Big] + \sum_{i=1}^N\sigma^\prime(\langle T^{i}\tilde{w}, s \rangle) \cdot \sum_{i=1}^N\sigma^\prime(\langle T^{i}\tilde{w}, s^\prime \rangle). \label{eq:decomp}
    \end{align}
    Recall that here $\tilde{w} \in \mathbb{R}^{N \times 2}$ has a support of size $K$. We use it in this way because it is applied to a full state $s$. Then, if we consider each motif of size $K$, $s_{i:i+K-1}$, we only need a $\tilde{w} \in \mathbb{R}^{K \times 2}$. Thus, in what follows, we slightly abuse $\tilde{w}$ as a $K \times 2$ kernel when applied to motifs. We can rewrite the second term in Eq.~\eqref{eq:decomp} as
    \begin{align}
        - N \sum_{i=1}^N \Big[\sigma^\prime(\langle \tilde{w}, s_{i:i+K-1} \rangle) + \sigma^\prime(\langle \tilde{w}, s_{i:i+K-1}^\prime \rangle)\Big], \label{eq:K_2}
    \end{align}
    Since $s$ and $s^\prime$ are arbitrary states, we only need to consider one of them. Then, we claim that $\forall s$, $\sum_{i=1}^N \sigma^\prime(\langle \tilde{w}, s_{i:i+K-1} \rangle) = N / 2$.
    
    To see this, we first show that there exists a partition of all $2^K$ possible motifs into two equal-sized sets $\mathcal{M}_{+}$ and $\mathcal{M}_{-}$, \textit{s.t.} $\forall m \in \mathcal{M}_{+}$, $\langle \tilde{w}, m \rangle \geq 0$ and $\forall m \in \mathcal{M}_{-}$, $\langle \tilde{w}, m \rangle \leq 0$. We will resolve the issue that both sets include the case $\langle \tilde{w}, m \rangle = 0$ later. This can be done since $\forall m \in \mathcal{M}$, we can always find its relabeling version $Lm$. When the grand sum condition holds, according to Eq.~\eqref{eq:gs_Ls}, we have $\langle \tilde{w}, m \rangle + \langle \tilde{w}, Lm \rangle = \langle \tilde{w}, m \rangle - \langle \tilde{w}, m \rangle = 0$. Hence, we can just put $m$ into $\mathcal{M}_{+}$ and $Lm$ into $\mathcal{M}_{-}$ if $\langle \tilde{w}, m \rangle > 0$; and vice versa. If $\langle \tilde{w}, m \rangle = 0$, we can just put $m$ into any one of the two sets and put $Lm$ into the other one.
    
    Next, consider each motif in the summation $\sum_{i=1}^N \langle \tilde{w}, s_{i:i+K-1} \rangle = 0$, according to Eq.~\eqref{eq:s} (a). For each $i$, if $s_{i:i+K-1} \in \mathcal{M}_{-}$, we replace $\langle \tilde{w}, s_{i:i+K-1} \rangle$ with $-\langle \tilde{w}, Ls_{i:i+K-1} \rangle$. Then, we have
    \begin{align*}
        \sum_{i=1}^N \langle \tilde{w}, s_{i:i+K-1} \rangle = \sum_{i \in [N] \cap \{i \mid s_{i:i+K-1} \in \mathcal{M}_{+}\}} \langle \tilde{w}, s_{i:i+K-1} \rangle - \sum_{j \in [N] \cap \{j \mid s_{j:j+K-1} \in \mathcal{M}_{-}\}} \langle \tilde{w}, Ls_{j:j+K-1} \rangle. 
    \end{align*}
    In cases where there are $i, j \in [N]$, \textit{s.t.} $s_{i:i+K-1} \in \mathcal{M}_{+}$, $s_{j:j+K-1} \in \mathcal{M}_{-}$ and $Ls_{j:j+K-1} = s_{i:i+K-1}$, these terms are canceled out. Let $\mathcal{I}$ denote the set of indices that remain after this canceling and $\forall i \in \mathcal{I}$, $s_{i:i+K-1} \in \mathcal{M}_{+}$. Let $\mathcal{J}$ denote the set of remaining indices such that $\forall j \in \mathcal{J}$, $s_{j:j+K-1} \in \mathcal{M}_{-}$ and that for all $k$ satisfying $s_{k:k+K-1} \in \mathcal{M}_{+}$, $Ls_{j:j+K-1} \ne s_{k:k+K-1}$. Then, the remaining summations become
    \begin{align}
        &\sum_{i \in \mathcal{I}} \langle \tilde{w}, s_{i:i+K-1} \rangle - \sum_{j \in \mathcal{J}} \langle \tilde{w}, Ls_{j:j+K-1} \rangle = \bigg\langle \tilde{w},  \sum_{i \in \mathcal{I}} s_{i:i+K-1} - \sum_{j \in \mathcal{J}} Ls_{j:j+K-1} \bigg\rangle = 0. \label{eq:symbol_counting}
    \end{align}
    
    We claim that $|\mathcal{I}| = |\mathcal{J}|$ for any non-trivial $\tilde{w}$. Suppose for contradiction, $|\mathcal{I}| \ne |\mathcal{J}|$. Recall that $s \in \mathbb{R}^{N\times 2}$. Each of its rows is a one-hot vector representing its spin. Let $S_{\mathcal{I}} = \sum_{i \in \mathcal{I}} s_{i:i+K-1}$ and $S_{\mathcal{J}} = \sum_{j \in \mathcal{J}} s_{j:j+K-1}$. Then, for the $i$th row of $S_{\mathcal{I}}$, the sum of the two elements $S_{\mathcal{I}, i, 1} + S_{\mathcal{I}, i, 2} = |\mathcal{I}|$. Similarly, for the $i$th row of $S_{\mathcal{J}}$, we have $S_{\mathcal{J}, i, 1} + S_{\mathcal{J}, i, 2} = |\mathcal{J}|$. Since, $|\mathcal{I}| \ne |\mathcal{J}|$, there is at least one non-zero value in each row of $S_{\mathcal{I}} - S_{\mathcal{J}}$. Hence, for Eq.~\eqref{eq:symbol_counting} to hold, i.e. $\langle \tilde{w}, S_{\mathcal{I}} - S_{\mathcal{J}} \rangle = 0$, for arbitrary $s$, we need to have $\tilde{w} = 0$, since any element in each row might be non-zero. This contradicts our assumption that $\tilde{w}$ is non-trivial.

    Therefore, when $\mathcal{I}$ and $\mathcal{J}$ are combined with the indices canceled out, we have that the size of the set $[N] \cap \{i \mid s_{i:i+K-1} \in \mathcal{M}_{+}\}$ equals the size of the set $[N] \cap \{j \mid s_{j:j+K-1} \in \mathcal{M}_{-}\}$. In other words, there are equal number of positive and negative $\langle \tilde{w}, s_{i:i+K-1} \rangle$, for $i=1,\cdots,2$. When fed into $\sigma^\prime(\cdot)$, they become an equal number of 0s and 1s. Therefore, $\forall s$, $\sum_{i=1}^N \sigma^\prime(\langle \tilde{w}, s_{i:i+K-1} \rangle) = N / 2$. And Eq.~\eqref{eq:K_2} becomes
    \begin{align*}
        -N \Big[\sum_{i=1}^N \sigma^\prime(\langle \tilde{w}, s_{i:i+K-1}) + \sum_{i=1}^N \sigma^\prime(\langle \tilde{w}, s_{i:i+K-1}^\prime)\Big] = -N \Big[\frac{N}{2} + \frac{N}{2}\Big] = -N^2.
    \end{align*}
    Finally, Eq.~\eqref{eq:decomp} becomes
    \begin{align*}
        &\sum_{i=1}^N \sigma^\prime(\langle \tilde{w}, T^{N-i}Ls \rangle) \cdot \sum_{i=1}^N \sigma^\prime(\langle \tilde{w}, T^{N-i}Ls^\prime \rangle) \nonumber \\
        =& N^2 - N \sum_{i=1}^N \Big[\sigma^\prime(\langle T^{i}\tilde{w}, s \rangle) + \sigma^\prime(\langle T^{i}\tilde{w}, s^\prime \rangle)\Big] + \sum_{i=1}^N\sigma^\prime(\langle T^{i}\tilde{w}, s \rangle) \cdot \sum_{i=1}^N\sigma^\prime(\langle T^{i}\tilde{w}, s^\prime \rangle) \nonumber \\
        =& N^2 - N^2 + \sum_{i=1}^N\sigma^\prime(\langle T^{i}\tilde{w}, s \rangle) \cdot \sum_{i=1}^N\sigma^\prime(\langle T^{i}\tilde{w}, s^\prime \rangle) \nonumber \\
        =& \sum_{i=1}^N\sigma^\prime(\langle T^{i}\tilde{w}, s \rangle) \cdot \sum_{i=1}^N\sigma^\prime(\langle T^{i}\tilde{w}, s^\prime \rangle).
    \end{align*}
    Thus, Eq.~\eqref{eq:K_w} is invariant to $g$ transformation.
    \item For $b$, we have
    \begin{align*}
        \partial_w\psi(s) = \psi(s) v \sum_{i=1}^N \sigma^\prime(\langle \tilde{w}, T^{N-i}s \rangle).
    \end{align*}
    Then, 
    \begin{align*}
        \partial_b^T\psi(s)\partial_b\psi(s^\prime) = \psi(s)\psi(s^\prime) v^2 \sum_{i=1}^N \sigma^\prime(\langle \tilde{w}, T^{N-i}s \rangle) \cdot \sum_{i=1}^N \sigma^\prime(\langle \tilde{w}, T^{N-i}s^\prime \rangle).
    \end{align*}
    Consider when applying $g$, we have
    \begin{align}
        \partial_b^T\psi(gs)\partial_b\psi(gs^\prime) = \psi(gs)\psi(gs^\prime) v^2 \sum_{i=1}^N \sigma^\prime(\langle \tilde{w}, T^{N-i}gs \rangle) \cdot \sum_{i=1}^N \sigma^\prime(\langle \tilde{w}, T^{N-i}gs^\prime \rangle). \label{eq:K_b}
    \end{align}
    All terms in Eq.~\eqref{eq:K_b} appear in Eq.~\eqref{eq:K_w}, which are all invariant to $g$. Therefore, Eq.~\eqref{eq:K_b} is invariant to $g$.
\end{enumerate}
Having considered the 3 cases above, we know that $\partial_\theta^T\psi(s)\partial_\theta\psi(s^\prime) = \partial_\theta^T\psi(gs)\partial_\theta\psi(gs^\prime)$ and hence the learning dynamics are invariant to $g$, when the grand sum condition holds. \qed

\section{Equivalence Classes and How to Count Them} \label{supp:ec}
If a general Hamiltonian $H$ acting on a Hilbert space $\mathcal{H}$ with a positive, nondegenerate ground state $\psi_{GS}$ possesses certain symmetries $\mathcal G$, then $\psi_{GS}$ will be invariant under $\mathcal G$. Therefore, since $\psi_{GS}(s)=\psi_{GS}(gs)$, there are only as many unique values of the wavefunction as there are equivalence classes of $\mathcal{H}$ under the action of $\mathcal G$. For the case of a 1D spin chain with translation, reflection, and permutation symmetries, the problem of calculating the exact number of equivalence classes is solved by de Bruijn's extension of Polya's enumeration theorem (Theorem 5.4 of \cite{deBruijn1964}), but it is inefficient to compute for large systems. However, a lower bound of this number can be easily obtained by noting that all equivalence classes have a maximum size equal to the total number of symmetries. This gives a lower bound of $$|\{\mathcal{E}\}|\geq\frac{\#\text{states}}{\#\text{symmetries}}$$ which is true for any system with a finite number of states. For the case of a 1D spin chain of $N$ particles with translation, reflection, and SU($M$) symmetry at equal concentrations, the lower bound is equal to $\frac{n!}{(\frac{n}{m})!^m}\frac{1}{m!2n}\sim O(M^N)$. For small systems $(N\sim10^1)$, the equivalence classes can be enumerated explicitly, at which point a reduced Hamiltonian can be generated, which mimics the full Hamiltonian but whose Hilbert space is the set of equivalence classes. This considerably speeds up any computation which accesses the Hamiltonian directly, such as exact diagonalization.

\section{Derivation of MaxEnt Ansatz with Symmetries} \label{suppsec:derive_maxent}
The MaxEnt formulation finds the best guess probability distribution subject to only information about the constraints. To adapt it to our wavefunction, we follow the method in \cite{Canosa1989} by exploiting the positive definiteness of our wavefunction to write $P(s)=\psi^2(s)$ as the probability of each string. Thus, we classical entropy is: 
\begin{equation}
    S=-\sum_s P(s)\ln P(s)=-\sum_s\psi^2(s)\ln\psi^2(s)
\end{equation} 
The constraints are some set of diagonal observables $\hat O_i$ are measured to have expectation values $O_i$ with respect to the ground state
\begin{equation}
    \langle \psi|\hat O_i|\psi\rangle=\sum_s P(s) O(s) = O_i\qquad \forall i\label{eq:constraints}
\end{equation}
In our case, we constrain the motif counting operators' ($m_{s'}(s)$) expectation values to match those of the true ground state:
\begin{equation}
    \sum_s P(s) m_{s'}(s)  = \langle m_{s'} \rangle_{GS}\qquad \forall s' \in \mathcal{Y}_{MK}
\end{equation}
We impose these constraints on the entropy using Lagrange multipliers $\lambda_{s'}$ for each $s'$:
\begin{equation}
    S'=-\sum_s P(s)\ln P(s)-\sum_{s'}\lambda_{s'}\left[\sum_s P(s) m_{s'}(s)-\langle m_{s'} \rangle_{GS}\right]
\end{equation}
To maximize, we functionally vary $S'$ w.r.t $P(s)$:
\begin{equation}
    \frac{\delta S}{\delta P(s)}
    =- \left[\ln P(s) +1-\sum_{s'}\lambda_{s'}m_{s'}(s)\right]=0
\end{equation}
This yields our MaxEnt ansatz as:
\begin{equation}
    P(s)=\psi^2(x)=\exp\left(-\sum_{s'}\lambda_{s'}m_{s'}(s)-1\right)
\end{equation}
\begin{equation}
    \psi^{\textrm{MaxEnt}}(x)=\frac{1}{Z}\exp\left(-\sum_{s'}\lambda_{s'}m_{s'}(s)\right)
\end{equation}
In general, there is a unique wavefunction of this form which satisfies Equation \eqref{eq:constraints}. However, we can also disregard the constraints and interpret this as a variational ansatz with parameters $\lambda_i$, which is the approach taken in the main text.

We can also reformulate MaxEnt over the equivalence classes of symmetries. Given a set of symmetries $\mathcal{G}$, we can always partition the Hilbert space into unions of symmetry equivalence classes $\{\mathcal{E}_k\}$.
Let the total number of equivalence classes be $\mathcal{N}$. Then, any expectation value of a \emph{diagonal} observable over the classical probability $P(s)$ may be rewritten as:
\begin{equation}
    \braket{\hat O}  =\sum_i^\mathcal{N} \left(\tilde P(\mathcal{E}_k) \sum_{s \in \mathcal{E}_k} O(s)\right)=\sum_k^\mathcal{N} \tilde P(\mathcal{E}_i) |\mathcal{E}_k|\tilde O(\mathcal{E}_k)
\end{equation}
where 
\begin{equation}
    \tilde O(\mathcal{E}_i)= \frac{1}{ |\mathcal{E}_i|}\sum_{s \in \mathcal{E}_i} O(s)
\end{equation}
In our case, we have
\begin{equation}
    \tilde m_{s'}(\mathcal{E}_k)= \frac{1}{ |\mathcal{E}_i|}\sum_{s \in \mathcal{E}_i}m_{s'}(s)
\end{equation}
Then we can rewrite:
\begin{equation}
    S=-\sum_k^\mathcal{N}|\mathcal{E}_k| \tilde P(\mathcal{E}_k) \ln{\tilde P(\mathcal{E}_k)}-\sum_{s'}\lambda_{s'}\left(\sum_k^{\mathcal{N}} \tilde P(\mathcal{E}_k)|\mathcal{E}_k|  \tilde m_{s'}(\mathcal{E}_k)- \braket{m_{s'}}_{GS}\right)
\end{equation}
We can now do functionally vary $S$  w.r.t  $\tilde P(\mathcal{E}_k)$ and set the result to zero:
\begin{equation}
    \frac{\delta S}{\delta P(\mathcal{E}_k)}=-\sum_{k}|\mathcal{E}_{k}|\left[\ln{\tilde \psi(\mathcal{E}_{k})^2}+1-\sum_{s'}\lambda_{s'} |\mathcal{E}_{k}| \tilde m_{s'}(\mathcal{E}_{k})\right]
\end{equation}
\begin{equation}
    \ln\psi^{\mathrm{MaxEnt}}(\mathcal{E}_k)=\sum_{s'}\lambda_{s'} \tilde m_{s'}(\mathcal{E}_k)+ \ln Z
\end{equation}
where $\ln Z$ is a renormalization term.

We can also write down a more restricted MaxEnt ansatz by considering motif equivalence classes. Denote the motif equivalence classes as $\{M_n\}$. Then for all motifs $s' \in M_n$, we have $\braket{m_{s'}}=\braket{m_n}$, where $\braket{m_n}$ is the MEV for motif class $M_n$. Furthermore, we can pick a representative motif $s'_n$ for each $M_n$:
\begin{equation}
    \braket{m_n}=\braket{m_{s'_n}}= \sum_s P(s) m_{s'_n}(s)
\end{equation}

Then our MaxEnt problem can be reformulated thus: maximize the classical entropy of a probability distribution $P(s) =|\psi(s)|^2$ s.t. :
\begin{equation}
    \braket{m_{s'_n}} = \braket{m_{s'_n}}_0 \quad \forall M_n
\end{equation}

Then our entropy functional:
\begin{align}
    S&= -\sum_s P(s) \ln P(s) +\sum_n \lambda_n \left[\braket{m_{s'_n}} - \braket{m_{s'_n}}_0 \right]\\
    S&= -\sum_s  \left[P(s) \ln P(s) -\sum_n \lambda_n \left[P(s)m_{s'_n}(s) - P_0(s)m_{s'_n}(s)\right] \right]\\
    \frac{\delta S}{\delta P(s)}&= -\sum_s \left[ \ln P(s) +1 -\sum_n \lambda_n m_{s'_n}(s)\right]=0
\end{align}
which leads us to the MaxEnt probability ansatz as
\begin{equation}
       P(s)=\frac{1}{Z}e^{\sum_n \lambda_n m_{s'_n}(s)}
\end{equation}
Thus, we can write the MaxEnt ansatz as one purely over the motif equivalence classes. This illustrates that the number of distinct non-zero Lagrange multipliers required to characterize the system are equal to or less than the number of motif equivalence classes.

\section{Entanglement Calculation Derivation and Errors} \label{supp:entanglement}
\begin{figure}
    \centering
    \includegraphics[width=.7\textwidth]{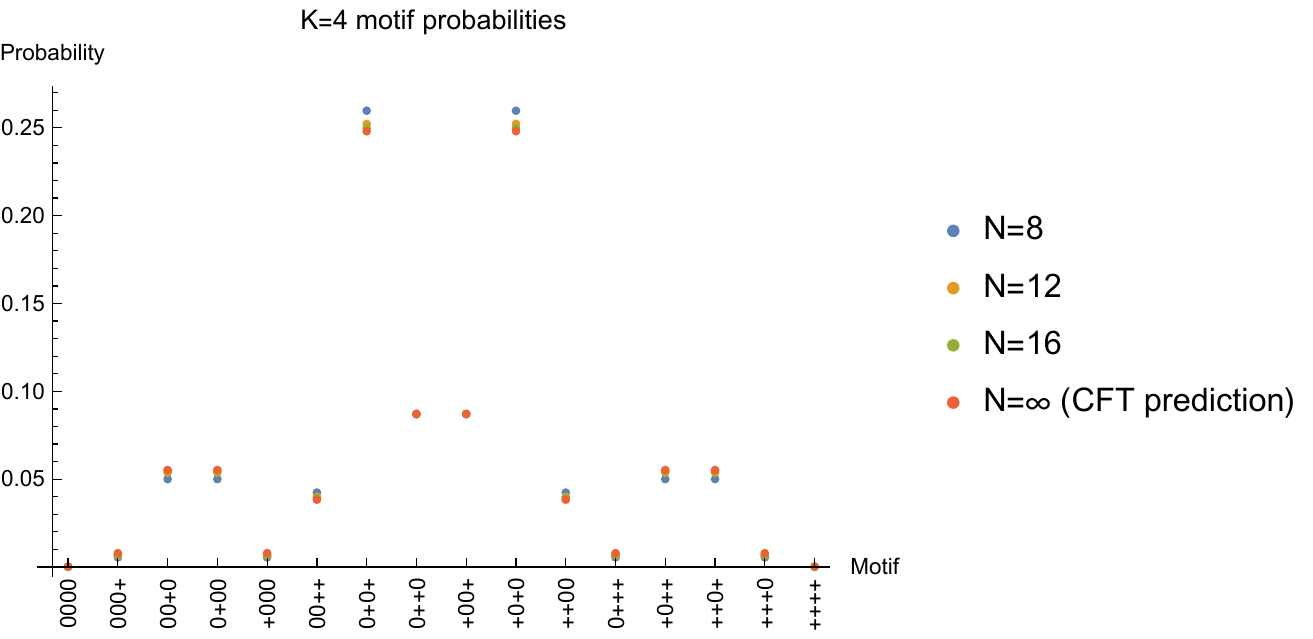}
    \caption{The $K=4$ motif probabilities for $N=\{8,12,16\}$ show a clear convergence to the values predicted by CFT in the thermodynamic limit. We take these to be our benchmarks for large $N$.}
    \label{fig:cftmotifs}
\end{figure}

The reduced density matrix of a group of locally connected particles has been of considerable interest over the last several decades, most notably for the DMRG algorithm \cite{white1992density}. In our case, it is useful to obtain density matrices for the purpose of calculating the exact MEVs without first computing the full exact GSWF. To do this, we employ results from the theory of entanglement Hamiltonians, which treats the subsystem as though it were immersed in a thermal bath. In this framework, the logarithm of the density matrix, called the entanglement Hamiltonian, has the same terms of the original Hamiltonian, but with position-dependent coefficient which scale linearly with the distance to the boundary \cite{cardy2016entanglement, giudici2018entanglement}. This has been analytically shown for various 1D systems such as the Ising \cite{peschel1999density} and free fermion \cite{eisler2017analytical} models. Here, we use the adaptation from \cite{Mendes-Santos2019}, which approximates the entanglement Hamiltonian as

\begin{equation}
    \rho_K=e^{-\beta H_K},\qquad H_K=\sum_{i=1}^{K-1}\frac{i(K-i)}{K}P_{i,i+1}
\end{equation}
Where $\beta$ is the effective inverse temperature related to the underlying conformal field theory \cite{cardy2016entanglement}. The errors coming from finite-size and lattice geometry effects are nonzero, but still small enough to be neglected for our purposes, as shown in Fig.~\ref{fig:cftmotifs}. Therefore, we take these values to be our benchmarks for large N experiments.

\section{Regression Analysis} \label{suppsec:reg_all}
 \begin{table}[h]
    \begin{center}
        \caption{Regression results for $\delta_E$ vs. $\bar{\delta}_d$, $d \in \{0, 1\}$, all 160 observations.}
        \begin{tabular}{rrrrr}
            \toprule
                    \textbf{Algorithm} &       \textbf{Original} &  \textbf{SymForce-Init} & \textbf{SymForce-Traj} &            \textbf{CPS} \\
            \midrule
     $R^2$ &                 0.891 &                 0.776 &                 0.831 &                 0.479 \\
  No. Obs. &                    40 &                    40 &                    40 &                    40 \\
 Cond. No. &                  18.4 &                  27.1 &                  26.5 &                  38.3 \\
 Intercept &    $1.648$* $(0.636)$ &    $1.655$* $(0.668)$ &   $2.103$** $(0.625)$ &   $0.811$** $(0.255)$ \\
$\delta_0$ & $-1.173$*** $(0.258)$ &    $-0.378$ $(0.426)$ &    $-0.408$ $(0.417)$ & $-1.043$*** $(0.284)$ \\
$\delta_1$ & $-8.469$*** $(0.520)$ & $-7.801$*** $(0.867)$ & $-8.141$*** $(0.825)$ & $-1.762$*** $(0.472)$ \\
       $K$ &    $-0.155$ $(0.125)$ &    $-0.254$ $(0.136)$ &   $-0.296$* $(0.128)$ &    $-0.086$ $(0.051)$ \\
            \bottomrule
        \end{tabular}
        \label{tab:regression_all_outliers}
    \end{center}
    Standard errors are reported in parentheses after the coefficients. * indicates significance at the 95\% level. ** indicates significance at the 99\% level. *** indicates significance at the 99.9\% level.
\end{table}

 The original data have 160 observations in total, where we vary the algorithm, the CNN kernel size $K$, the number of sample reuse $n_{\mathrm{opt}}$ and the learning rate $\eta$. We run each hyperparameter combination using 5 random initializations (see Sec. \ref{suppsec:hyperparameters_bar} for details). For the analysis in Table \ref{tab:regression_all}, we estimate the energy gap $E_1-E_0$ as $\frac{1}{\mathcal{N}}\left(E_\mathrm{max}-E_0\right)$, where $E_\mathrm{max}$ is the largest energy eigenvalue and $\mathcal{N}$ is the size of the Hilbert space. We remove the observations with $\delta_E \geq 6$ since these do not learn the wavefunction properly and may greatly bias the regression results. If these outlier are included, as we can see in Table \ref{tab:regression_all_outliers}, indeed, the $R^2$s for the 3 CNN models become very high, since it is much easier to distinguish outliers from normal data points than model the finer-level structure within the normal data points. However, in either case, our arguments regarding the coefficient of $K$ holds.

\section{Hyperparameters and Tuning}
\subsection{For Results Shown in Fig.~\ref{fig:E-hat-alg}} \label{suppsec:hyperparameters}
We run the experiments for $N = 60$ and $N = 240$. We draw 1,000 samples in each iteration of variation Monte Carlo and train the models for 500 iterations. We also use the following hyperparameters: the kernel size $K \in \{3, 6, 12, 24\}$, the learning rate $\eta \in \{0.0001, 0.001\}$, the number of iterations reusing the same VMC samples during training $n_{\mathrm{opt}} \in \{10, 100\}$, and the training algorithms $\mathcal{A} \in$ $\{$Original, Deep ($L$ layers), SymForce-Init, SymForce-Traj$\}$. For $\mathcal{A} \in$ $\{$Original, SymForce-Init, SymForce-Traj$\}$, we use $L = F = 1$. And when $\mathcal{A} = \text{Deep (}L{\text{ layers)}}$, we follow \cite{yang_deep_2020} for the choices of $F$ and $L$. We use $L \in \{2, 3, 8\}$. For 2-layer CNNs, we use $F \in \{1, 8, 16\}$ and for deeper CNNs, we use $F \in \{8, 16\}$. For hyperparameter tuning, we run each setting 5 times with different random initializations by setting the random seed of TensorFlow \cite{abadi2016tensorflow} and NumPy \cite{van2011numpy}. After we collect the experiment results, we first remove the hyperparameter settings causing divergence in any of the 5 runs. Then, for each $(\mathcal{A}, L, K)$, we pick the hyperparameters that lead to the minimum absolute value of the relative error in the ground state energy.
\subsection{For Results Shown in Fig.~\ref{fig:mev_all}, Table \ref{tab:regression_all} and Table \ref{tab:regression_all_outliers}} \label{suppsec:hyperparameters_bar}
We use the following hyperparameters for both shallow CNN and CPS models trained for $N=60$ systems: $K \in \{3, 6\}$, $\eta \in \{0.0001, 0.001\}$, $n_{\mathrm{opt}} \in \{10, 100\}$, $\mathcal{A} \in$ \{\texttt{original}, \texttt{grand} \texttt{-} \texttt{sum} \texttt{-} \texttt{init}, \texttt{grand} \texttt{-} \texttt{sum} \texttt{-force}$, \texttt{CPS}\}$. We also list the hyperparameters that achieve the least error for each algorithm: 
\begin{enumerate}
    \item Original: $K=6$, $\eta = 0.001$, $n_{\mathrm{opt}} = 100$, $\mathrm{seed} = 1$.
    \item SymForce-Init: $K=6$, $\eta = 0.001$, $n_{\mathrm{opt}} = 10$, $\mathrm{seed} = 3$.
    \item SymForce-Traj: $K=6$, $\eta = 0.001$, $n_{\mathrm{opt}} = 10$, $\mathrm{seed} = 4$.
    \item CPS: $K=3$, $\eta = 0.0001$, $n_{\mathrm{opt}} = 100$, $\mathrm{seed} = 3$.
\end{enumerate}

\newpage

\bibliography{scibib}

\begin{thebibliography}{10}

\bibitem{Liang2021}
X.~Liang, S.-J. Dong, L.~He, {\it Phys. Rev. B\/} {\bf 103}, 035138 (2021).

\bibitem{miles2020correlator}
C.~Miles, {\it et~al.\/}, Correlator convolutional neural networks: An
  interpretable architecture for image-like quantum matter data (2020).

\bibitem{roth2021group}
C.~Roth, A.~H. MacDonald, {\it arXiv preprint arXiv:2104.05085\/}  (2021).

\bibitem{liang2018solving}
X.~Liang, {\it et~al.\/}, {\it Physical Review B\/} {\bf 98}, 104426 (2018).

\bibitem{Sutherland1975}
B.~Sutherland, {\it Phys. Rev. B\/} {\bf 12}, 3795 (1975).

\bibitem{yang_deep_2020}
L.~Yang, {\it et~al.\/}, {\it Physical Review Research\/} {\bf 2}, 012039
  (2020). Publisher: American Physical Society.

\bibitem{deng2017quantum}
D.-L. Deng, X.~Li, S.~D. Sarma, {\it Physical Review X\/} {\bf 7}, 021021
  (2017).

\bibitem{harney2020entanglement}
C.~Harney, S.~Pirandola, A.~Ferraro, M.~Paternostro, {\it New Journal of
  Physics\/} {\bf 22}, 045001 (2020).

\bibitem{sun2022entanglement}
X.-Q. Sun, T.~Nebabu, X.~Han, M.~O. Flynn, X.-L. Qi, {\it arXiv preprint
  arXiv:2203.00020\/}  (2022).

\bibitem{li2008entanglement}
H.~Li, F.~D.~M. Haldane, {\it Physical review letters\/} {\bf 101}, 010504
  (2008).

\bibitem{peschel1999density}
I.~Peschel, M.~Kaulke, {\"O}.~Legeza, {\it Annalen der Physik\/} {\bf 8}, 153
  (1999).

\bibitem{perez2006matrix}
D.~Perez-Garcia, F.~Verstraete, M.~M. Wolf, J.~I. Cirac, {\it arXiv preprint
  quant-ph/0608197\/}  (2006).

\bibitem{klumper1993matrix}
A.~Kl{\"u}mper, A.~Schadschneider, J.~Zittartz, {\it EPL (Europhysics
  Letters)\/} {\bf 24}, 293 (1993).

\bibitem{Mendes-Santos2019}
T.~Mendes-Santos, G.~Giudici, M.~Dalmonte, M.~A. Rajabpour, {\it Phys. Rev.
  B\/} {\bf 100}, 155122 (2019).

\bibitem{hermann2020deep}
J.~Hermann, Z.~Sch{\"a}tzle, F.~No{\'e}, {\it Nature Chemistry\/} {\bf 12}, 891
  (2020).

\bibitem{spencer2021learning}
J.~Spencer, {\it Nature Reviews Physics\/} {\bf 3}, 458 (2021).

\bibitem{Canosa1989}
N.~Canosa, A.~Plastino, R.~Rossignoli, {\it Phys. Rev. A\/} {\bf 40}, 519
  (1989).

\bibitem{losada2019solutions}
M.~Losada, F.~Holik, C.~Massri, A.~Plastino, {\it Quantum Information
  Processing\/} {\bf 18}, 1 (2019).

\bibitem{strang2016introduction}
G.~Strang, {\it Introduction to linear algebra 5th Edition\/}
  (Wellesley-Cambridge Press Wellesley, MA, 2016).

\bibitem{jacot2018neural}
A.~Jacot, F.~Gabriel, C.~Hongler, {\it Advances in neural information
  processing systems\/} {\bf 31} (2018).

\bibitem{deBruijn1964}
N.~G. De~Bruijn, {\it Applied combinatorical mathematics\/} (1964), pp.
  144--184.

\bibitem{white1992density}
S.~R. White, {\it Physical review letters\/} {\bf 69}, 2863 (1992).

\bibitem{cardy2016entanglement}
J.~Cardy, E.~Tonni, {\it Journal of Statistical Mechanics: Theory and
  Experiment\/} {\bf 2016}, 123103 (2016).

\bibitem{giudici2018entanglement}
G.~Giudici, T.~Mendes-Santos, P.~Calabrese, M.~Dalmonte, {\it Physical Review
  B\/} {\bf 98}, 134403 (2018).

\bibitem{eisler2017analytical}
V.~Eisler, I.~Peschel, {\it Journal of Physics A: Mathematical and
  Theoretical\/} {\bf 50}, 284003 (2017).

\bibitem{abadi2016tensorflow}
M.~Abadi, {\it et~al.\/}, {\it 12th USENIX symposium on operating systems
  design and implementation (OSDI 16)\/} (2016), pp. 265--283.

\bibitem{van2011numpy}
S.~Van Der~Walt, S.~C. Colbert, G.~Varoquaux, {\it Computing in science \&
  engineering\/} {\bf 13}, 22 (2011).

\end{thebibliography}
\bibliographystyle{Science}

\end{document}